# Magma Buoyancy Couples Lunar Mantle Convection and Magmatism: Long-Lasting Mare Volcanism and Radius Change in a 3-D Regional Model

**Ken'yo U[1,2*], Masanori Kameyama[1,3], Takatoshi Yanagisawa[3,4], Takehiro Miyagoshi[3], and Masaki Ogawa[2†]**

[1] Geodynamics Research Center (GRC), Premier Institute for Advanced Studies (PIAS), Ehime University, Matsuyama, Ehime 790-8577, Japan.

[2] Department of Earth Sciences and Astronomy, The University of Tokyo at Komaba, Meguro, Tokyo 153-8902, Japan.

[3] Japan Agency for Marine-Earth Science and Technology (JAMSTEC), Yokohama, Kanagawa 236-0001, Japan.

[4] Faculty of Engineering, Hokkaido University, Sapporo, Hokkaido 060-8628, Japan.

Corresponding author: Ken'yo U (kenyo.u.grc@gmail.com)

*"U" is the family name.

†(retired)

**Key Points:**

- We developed a 3-D regional spherical shell model of mantle convection and magmatism, showing that melt buoyancy in plumes drives convection
- Plumes repeatedly rise through thermally softened conduits formed by earlier plumes and sustain localized volcanism for billions of years
- Magma ascent by the plumes, which peaks between 4 and 3.5 Gyr ago, accounts for the observed mare volcanism and early expansion of the Moon

## Abstract

To understand the mantle evolution that has caused the observed histories of mare volcanism and radius change of the Moon, we developed a 3-D regional spherical shell mantle convection model with magmatism. Magma generated by decompression melting segregates from the convecting matrix and percolates upward through it. In our model, internal heating forms a basal partially molten layer during the first several hundred million years. Finger-like magma ascents then grow from the top of this layer and evolve into buoyant, partially molten plumes. The plumes ascend by the buoyancy of magma, while decompression melting in the ascending plumes generates more magma that further drives the ascent. Magma supply to the top of the mantle peaks between 4 and 3.5 Gyr ago, coincident with a maximum in radial expansion driven by volume change of the mantle associated with melting, and both decline as the mantle cools and solidifies. Magma supply, however, persists for billions of years because earlier plumes leave thermally softened conduits, through which later plumes repeatedly rise at fixed locations. These stable conduits are a consequence of the 3-D geometry. Recycling of magmatic products formed in the uppermost mantle also triggers renewed magma ascent through the conduits. Our results provide a mechanism for the observed long-lasting localized volcanism that peaked at 4-3.5 Gyr ago and quantitatively account for early expansion followed by contraction. The conditions required for consistency with the observed history imply an initially hot but mostly solid mantle with a dense basal layer.

## Plain Language Summary

We developed a three-dimensional regional spherical shell mantle convection model with magmatism to understand the observed history of lunar mare volcanism as well as radius change, with emphasis on the long-lasting, localized volcanism and the early radial expansion of the Moon. In our numerical models, internal heating first forms a basal partially molten layer. Then, finger-like magma upwellings grow from the top of this layer and evolve into buoyant, partially molten plumes. The plumes rise due to the buoyancy of the melt, while melting in the rising plumes generates more magma, further driving their ascent. The extension of partially molten regions contributes to radial expansion owing to volume change associated with mantle melting, while the plumes cause volcanism. The volcanism continues for billions of years at fixed locations because earlier plumes leave (i) thermally softened conduits, through which later magma repeatedly rises, and (ii) magmatic products that later sink and induce counterflows, which trigger renewed magma ascent through the conduits. This mantle evolution accounts for both observed histories of volcanism and radius change; the model also implies that the early mantle was hot but mostly solid, with a dense basal layer enriched in heat-producing elements.

## 1 Introduction

Understanding how the lunar mantle has evolved to produce the observed history of mare volcanism and global radius change has been a long-standing issue in studies of the interiors of terrestrial planets (for a review, see Breuer & Moore, 2015). Geological and remote-sensing observations show that mare volcanism began within the first several hundred million years of lunar history and reached a peak at around 3.8–3.5 Gyr ago (e.g., Hiesinger et al., 2000, 2003; Morota et al., 2011a; Hurwitz et al., 2013; Jin et al., 2025). The thickest mare deposits and the majority of eruptive centers are concentrated on the lunar nearside, particularly within and around the Procellarum KREEP Terrane (PKT) (e.g., Jolliff et al., 2000; Wieczorek & Phillips, 2000; Giguere et al., 2022). After its peak, the volcanism declined with time but persisted until around 1.5 Gyr ago (e.g., Hiesinger et al., 2000; Morota et al., 2011b; Whitten & Head, 2015a). This prolonged and temporally variable volcanic activity broadly corresponds to the history of radial expansion and contraction of the Moon: gravity gradiometry data indicate that the Moon experienced an early phase of global expansion of about 0.6–4.9 km that lasted until around 3.8 Gyr ago (e.g., Andrews-Hanna et al., 2013, 2014; Sawada et al., 2016; Liang & Andrews-Hanna, 2022; Nishiyama et al., 2024). After this expansion, the Moon globally contracted to the present, as inferred from tectonic features such as lobate scarps and wrinkle ridges, with estimates suggesting that contraction over the past 1 Gyr is less than about 1 km (e.g., Watters et al., 2010, 2015, 2026; Clark et al., 2017; Yue et al., 2017; Matsuyama et al., 2021; Frueh et al., 2023). To clarify the mantle evolution responsible for these observed features of lunar history, we developed a three-dimensional (3-D) regional spherical shell model of lunar mantle evolution in which both mantle convection and magmatism transport heat, mass, and incompatible heat-producing elements (HPEs).

Previously, spherically symmetric one-dimensional (1-D) numerical models of the thermal evolution of the mantle driven by internal heating and conductive heat transport have been proposed to account for the observed radius change (e.g., Lyubimova, 1977; Solomon, 1977, 1986). Classical 1-D models, where the radius change occurs solely thermally, show that early expansion that is consistent with the observed one occurs when the temperature in the deep mantle is initially 1100 K or less (e.g., Solomon & Head, 1979); viewed in the light of the early expansion inferred today, these classical models imply that the deep mantle must have started cold and must then have been heated by more than 700 K through internal heating (Solomon & Chaiken, 1976). Even when additional volume changes associated with compositional differentiation are considered, the range of initial deep-mantle temperatures that permits early expansion is only modestly broadened to around 1200 K (Kirk & Stevenson, 1989). Giant-impact hypotheses for the origin of the Moon, however, suggest a much higher initial thermal state; the mantle is expected to have been largely molten, forming a deep magma ocean after the impact (e.g., Stevenson, 1987; Pritchard & Stevenson, 2000; Canup, 2004; Lock et al., 2018; Hosono et al., 2019; Kegerreis et al., 2022). Even at the end of mantle overturn, which is expected to occur after solidification of the magma ocean (see Section 2), the temperature of the deep mantle is expected to remain as high as ~1800–1900 K (e.g., Hess & Parmentier, 1995, 2001; Alley & Parmentier, 1998; Boukaré et al., 2018; H. Li et al., 2019). To address this discrepancy, U et al. (2022) constructed a 1-D spherically symmetric model with a high initial temperature that explicitly considers mantle melting and the migration of the generated magma by melt percolation, and examined the volume change of the mantle associated with melting. In this model, internal heating forms a partially molten region in the deep mantle, and the generated magma percolates upward and extends the region. The extended partially molten region persists for billions of years, and its volume change by melting contributes to the early expansion in a manner consistent with the observations. This result suggests that magma generation and migration played an important role in the early expansion of the Moon. Yet, implications for volcanic activity are limited because mantle convection is not considered in this model.

To clarify how the heat and mass transport by mantle convection can exert control over the evolution of the lunar mantle, various two-dimensional (2-D) and 3-D numerical models have been developed (for a review, see Tosi & Padovan, 2021). Some earlier models suggest that mantle convection can keep the uppermost mantle super-solidus to cause volcanic activity for billions of years if the surface is covered with a crust enriched in HPEs or a blanketing regolith layer (Konrad & Spohn, 1997; Spohn et al., 2001; Ziethe et al., 2009). Some other models assume that a part of the uppermost mantle on the nearside is more enriched in HPEs than that on the farside as inferred from the PKT (Pieters et al., 1980; Wieczorek & Phillips, 2000) and suggest that the enriched part of the mantle can remain super-solidus for more than 3 billion years (Laneuville et al., 2013, 2014, 2018). How the mantle differentiation induced by the lunar magma ocean at the beginning of lunar history has influenced the subsequent history of the thermal state of the mantle has also been a subject of active research. These studies suggest that a compositionally dense layer enriched in ilmenite-bearing cumulates (IBC) and HPEs developed at the base of the mantle during crystallization of the magma ocean (H. Li et al., 2019; Yu et al., 2019; Maurice et al., 2024; see also Section 2 below) and rose as one or more plumes due to thermal buoyancy from its own internal heating; this plume activity has been used to explain the timing and regional distribution of volcanic activity of the Moon (e.g., Zhong et al., 2000; Stegman et al., 2003; N. Zhang et al., 2013a, b; W. Zhang et al., 2022, 2023). To link mantle dynamics to volcanism more directly, however, it is necessary to model magma generation and ascent numerically, rather than relying on discussions based on the spatial distribution and temporal evolution of the super-solidus region (e.g., Cassen & Reynolds, 1973; Cassen et al., 1979; Ogawa, 2014, 2018a; Dannberg & Heister, 2016). Recently, we constructed a 2-D model of mantle convection and magmatism for the Moon to infer volcanic history directly from the calculated history of mantle melting (U et al., 2023, 2025). These studies show that magma generated in the deep mantle can ascend to the base of the crust as partially molten plumes and that this process, coupled with the extension of partially molten regions, can account for the observed long-lasting volcanic activity and radial expansion. However, some studies show that mantle dynamics and subsequent plume activity can critically depend on the geometry of numerical models (e.g., Hernlund & Tackley, 2008; Kameyama, 2021; R. Zhang et al., 2025); the differences between 2-D polar-rectangular and 3-D spherical models are particularly significant when the core size is small, as is the case for the Moon (Guerrero et al., 2018; Fleury et al., 2024). It is therefore necessary to verify whether the mechanisms demonstrated in 2-D models are also valid for 3-D geometries and to perform a quantitative assessment by comparing the model results with observations.

To understand the mantle evolution of the Moon as constrained by the volcanic history and the radial expansion/contraction history, here, we extend the 2-D polar-rectangular model of a coupled magmatism-mantle convection system we developed earlier (U et al., 2023, 2025) to a 3-D regional spherical shell model. In this study, we simulate 4.4 Gyr of mantle evolution driven by mantle convection, magma generation by decompression melting and internal heating, and upward percolation of the generated magma through the matrix. We start the calculation from a thermochemical state of the mantle predicted by earlier models of the magma ocean and mantle overturn that are believed to have occurred around 4.4 Gyr ago (e.g., Hess & Parmentier, 1995; Elkins-Tanton et al., 2011; Boukaré et al., 2018; Morbidelli et al., 2018; Maurice et al., 2020, 2024). We then compare the history of upward flux of magma into the uppermost mantle and the volume change of the mantle due to melting and thermal expansion, which we can calculate directly from our numerical model, with the observations. Although our calculations adopt a 3-D spherical shell geometry, they represent a regional model tailored to isolate how 3-D geometry influences mantle convection and magmatism; the model does not explicitly incorporate the global-scale lateral variations in crustal thickness, HPEs, and surface structure associated with the nearside-farside dichotomy and the distinctive characteristics of the PKT (e.g., Lawrence et al., 2000; Wieczorek & Phillips, 2000; Arai et al., 2008; Ohtake et al., 2012; Guo et al., 2024; Z. Li et al., 2026a, b).

## 2 Model Descriptions

We model the evolution of the lunar mantle driven by magmatism and mantle convection in a 3-D regional spherical shell $R = \left[(r, \theta, \psi) | \; r_c \leq \; r \leq r_p, \; \pi/4 \leq \theta \leq 3\pi/4, \; 0 \leq \; \psi \leq \pi\right]$ that extends from the core-mantle boundary (CMB) at $r_c = 385$ km to the surface at $r_p = 1735$ km, where $r$, $\theta$, and $\psi$ are the radius, colatitude, and longitude, respectively. The mantle material is a binary eutectic system denoted as $A_\xi B_{1-\xi}$ where $A$ is an olivine-rich end-member and $B$ is an IBC end-member; its eutectic composition at $\xi = \xi_e$ corresponds to a basalt enriched in the IBC component (Section 2.1). The IBC end-member is denser than the olivine-rich end-member. When a magma of the eutectic composition segregates from a mantle material with $\xi > \xi_e$, therefore, the residue of magma is less dense than the original solid material. The mantle contains incompatible HPEs that decay with time. The density of the mantle material depends on temperature, composition, and melt content. The mantle behaves as a highly viscous Newtonian fluid whose viscosity strongly depends on temperature, and its convection is driven by thermal, compositional, and melt buoyancy. Magma is generated by decompression melting and internal heating, and its composition is close to the eutectic composition in most cases. The generated magma is less dense than the coexisting matrix in most cases and migrates upward through the matrix as a permeable flow driven by its own buoyancy, carrying the IBC component and HPEs with it. To model the blanket effect of the crust (Ziethe et al., 2009), we reduced the thermal diffusivity by a factor of two in a blanket layer 35 km thick at the top of the mantle (Wieczorek et al., 2013). We present the employed basic equations in Section 2.1 below.

As for the boundary condition, the temperature is fixed at $T_{sur} = 270$ K on the surface boundary, while the core is modeled as a heat bath of uniform temperature; the vertical sidewalls are insulating. All of the boundaries are impermeable for both magma and matrix and are shear stress-free. In the initial condition, we assume that the deep mantle is more enriched in the HPEs and the IBC component than the shallower mantle as earlier models of mantle differentiation during crystallization of the magma ocean and subsequent mantle overturn suggest (e.g., Ringwood & Kesson, 1976; Hess & Parmentier, 1995; Maurice et al., 2020, 2024; Moriarty et al., 2021): during the last stage of crystallization of the lunar magma ocean, a dense IBC layer enriched in KREEP (a component rich in potassium, rare earth elements, and phosphorus) develops at the top of the mantle and subsequently sinks to the CMB, a process called mantle overturn (e.g., Snyder et al., 1992; Alley & Parmentier, 1998; de Vries et al., 2010; Elkins-Tanton et al., 2011; Boukaré et al., 2018). Several previous studies, however, suggest that a portion of the IBC material enriched in HPEs persists just beneath the crust even after mantle overturn (e.g., Yu et al., 2019; Zhao et al., 2019; Schwinger & Breuer, 2022). Instead of explicitly simulating these remains of HPE-enriched material, we assume that the crust is uniformly enriched in HPEs.

### 2.1 The Basic Equations

The densities of the olivine-rich end-member $A$ and the IBC end-member $B$ of the binary eutectic system $A_\xi B_{1-\xi}$ introduced above are 3300 kg m$^{-3}$ (Elkins-Tanton et al., 2011) and 3745 kg m$^{-3}$ (Snyder et al., 1992; Shearer et al., 2006; Elkins-Tanton et al., 2011; Rapp & Draper, 2018; Liang et al., 2024), respectively. The eutectic composition is $A_{0.1}B_{0.9}$ ($\xi = \xi_\mathrm{e} \equiv 0.1$). The density $\rho$ is written as

$$\rho = (1-\phi)\rho_\mathrm{s} + \phi\rho_\mathrm{l}\,, \tag{1}$$

where $\rho_\mathrm{s}$ is the density of solid phase, $\rho_\mathrm{l}$ the density of melt phase, and $\phi$ the melt content. The density depends on the temperature $T$ and the content of the end-member $A$ in the solid phase $\xi_\mathrm{s}$ and that in the melt phase $\xi_\mathrm{l}$ as:

$$\rho_\mathrm{s} = \rho_0[1-\alpha_m(T-T_\mathrm{sur}) + \beta(1-\xi_\mathrm{s})]\,, \tag{2}$$

and

$$\rho_\mathrm{l} = \rho_0\left\{1-\alpha_m(T-T_\mathrm{sur}) + \beta(1-\xi_\mathrm{l}) - \frac{\Delta V_\mathrm{l}}{V_0}[1+\beta(1-\xi_\mathrm{l})]\right\}\,, \tag{3}$$

where $\rho_0$ is the reference density; $\alpha_m$ the thermal expansivity; $\beta = 0.135$ a constant estimated from the density of olivine-rich end-member $A$ and that of the IBC end-member $B$; $T$ the temperature; $\Delta V_\mathrm{l}/V_0$ expresses the amount of density reduction by melting as:

$$\frac{\Delta V_\mathrm{l}}{V_0} = \frac{1}{\rho_0}\left[\Delta\rho_\infty + \frac{\Delta\rho_\mathrm{zero} - \Delta\rho_\infty}{(P/\lambda+1)^2}\right]\,. \tag{4}$$

Here, $P = -\int_{r_p}^{r}\rho_0\, g dr$ is the lithostatic pressure; $g = g_0 - (g_0 - g_\mathrm{r})\frac{r_p - r}{L}$ is the depth-dependent gravitational acceleration (Garcia et al., 2011, 2012); $L = r_\mathrm{p} - r_\mathrm{c}$ is the depth of the mantle. $\Delta\rho_\infty/\rho_0 = 0.005$ is the dimensionless density difference between solid and liquid phases at infinite pressure, and $\Delta\rho_\mathrm{zero}/\rho_0 = 0.22$ that at zero pressure. We assumed the value of $\lambda = 16.42$ GPa so that the solidus temperature defined below becomes close to that in the lunar mantle (Katz et al., 2003; Garcia et al., 2011, 2012).

The solidus temperature $T_{\text{solidus}}$ is calculated from the Clausius-Clapeyron relationship and depends on the pressure as:

$$T_{\text{solidus}} = T_0(1 + G)\,, \tag{5}$$

where

$$G = \frac{1}{\rho_0 \Lambda_0} \int_0^P \frac{\Delta V_l}{V_0}\, dP\,. \tag{6}$$

Here, $T_0$ is the solidus temperature at the surface (Katz et al., 2003); $\Lambda_0$ is the latent heat of melting. The solidus temperature calculated with the mentioned parameter values is shown in Figure 1a by the black dashed line. (The brown dashed line in Figure 1a is the solidus for Case inv-vl, where the parameter values in Eq. (4) are different; see Section 4.2.3.)

The continuity equation is

$$\nabla \cdot \mathbf{U} = -\nabla \cdot [\phi(\mathbf{u} - \mathbf{U})]\,, \tag{7}$$

where $\mathbf{U}$ and $\mathbf{u}$ stand for the velocity of matrix and magma, respectively.

The momentum equation for mantle convection is

$$-\nabla p - \rho g_0 \mathbf{e_r} + \nabla \cdot [\eta(\nabla\mathbf{U} + {}^t\nabla\mathbf{U})\,] = 0\,, \tag{8}$$

where $p$ is the dynamic pressure; $\mathbf{e_r}$ is the unit vector in the radial direction positive upward; the superscript $t$ means transpose of matrix. In Eq. (8), the gravitational acceleration is fixed at its surface value $g_0$ for simplicity. We confirmed that replacing $g_0$ in Eq. (8) with the depth-dependent $g$ does not affect the overall features of mantle evolution and the history of radius change and magmatism presented below (see Text S1, Figure S1, and Movie S7 in the Supporting Information). The viscosity of the mantle $\eta$ depends on the temperature as:

$$\eta = \eta_0 \exp[E_T(T_{\text{ref}} - T)]\,, \tag{9}$$

where $\eta_0$ is the reference viscosity; $T_{\text{ref}} = 1575\ \text{K}$ is the reference temperature; $E_T$ is the sensitivity of viscosity to temperature. The adopted default value of $E_T$ is $11.3 \times 10^{-3}\ \text{K}^{-1}$, which implies that the viscosity decreases by a factor of 3 as temperature increases by 100 K. This is an appropriate value to mimic a thermal convection of a non-Newtonian fluid with temperature-dependent viscosity by that of a Newtonian fluid (e.g., Christensen, 1984; Dumoulin et al., 1999). We adopted the Frank-Kamenetskii law rather than the Arrhenius law because the nature of the thermal convection with the former law has been thoroughly investigated (e.g., Yanagisawa et al., 2016). At the default value of $E_T$, the temperature dependence induces a viscosity contrast between the surface and deep mantle as large as 6 orders of magnitude, which is large enough for the lithosphere to develop as a stagnant lid when thermal convection takes place (Yanagisawa et al., 2016). The difference in viscosity between the Frank-Kamenetskii law and the Arrhenius law is significant only in the cold stagnant lid and does not have much effect on convection beneath the lid. We also confirmed that the overall features do not change much when $E_T$ is varied from half to 1.5 times its default value (Cases Et3 and Et9; Text S2 and Figure S2 in the Supporting Information).

In a partially molten region generated by decompression melting and internal heating, the difference between the velocity of magma $\mathbf{u}$ and that of matrix $\mathbf{U}$ is proportional to the density difference between them (Darcy's law):

$$\mathbf{u} - \mathbf{U} = \frac{k_\phi}{\phi \eta_{\mathrm{melt}}} g(\rho_s - \rho_l)\mathbf{e_r}\,, \tag{10}$$

where $k_\phi$ is the permeability that depends on the melt content $\phi$ as

$$k_\phi = k_{\phi_0} \left(\frac{\phi}{\phi_0}\right)^3 \tag{11}$$

(McKenzie, 1984). We assumed that $\phi_0 = 0.05$ and truncated $\phi$ at 0.4 for numerical reasons. The assumed value of $k_{\phi_0}$ given in Table 1 is based on the earlier works of McKenzie (1984) and Miller et al. (2014). Eq. (10) is a simplified form of Darcy's law for a deformable matrix (Eq. (A11) of McKenzie, 1984); we neglected the contribution from the compaction of the matrix because we are concerned with magma migration on the spatial scale of the mantle, which is much larger than the compaction length, around 1 km or less in the lunar mantle (McKenzie, 1984; U et al., 2026a). Our model therefore does not resolve the localized melt structures that develop on the scale of the compaction length, such as melt channels and solitary porosity waves (e.g., Scott & Stevenson, 1986; Spiegelman, 1993a, b; Aharonov et al., 1995; Keller & Katz, 2016). To our knowledge, no earlier study has shown that these small-scale structures grow into mantle-scale features; how they affect the mantle-scale magma migration treated here, however, remains to be studied. In the topmost 150 km of the spherical shell, we replaced $\phi$ in Eq. (11) with $\phi' = \max(\phi, \phi_e)$ to mimic magma migration that occurs by crack propagation in the crust and the uppermost mantle of the Moon (Head & Wilson, 2017; Wilson & Head, 2017). Eq. (11) then becomes

$$k_\phi = k_{\phi_0} \frac{[\max(\phi, \phi_e)]^3}{\phi_0^3}\,, \tag{12}$$

where $\phi_e = 0.035$ is an 'effective porosity' for magma migration by crack propagation (Kameyama et al., 1996).

The energy equation is

$$\frac{\partial(\rho_0 h)}{\partial t} + \nabla \cdot (\rho_0 h \mathbf{U}) = -\nabla \cdot [\rho_0 h_l \phi (\mathbf{u} - \mathbf{U})] - \frac{\Delta V_l}{V_0} \rho_0 g \phi u_r + \nabla \cdot (k \nabla T) + \rho_0 q + \nabla \cdot [\kappa_{\mathrm{edd}} \nabla (\rho_0 h)] , \tag{13}$$

where $h$ is the enthalpy defined by

$$h = C_p T + \phi \Delta h ; \tag{14}$$

$C_p$ is the specific heat capacity; $\Delta h = \Lambda_0 (1 + G)$; $h_l = h(\phi = 1)$; $q$ is the internal heating rate; $k = \rho_0 C_p \kappa$ is the thermal conductivity; $\kappa$ is the thermal diffusivity. We assumed that the matrix disintegrates and that a strong turbulent diffusion occurs with the eddy diffusivity of $\kappa_{\mathrm{edd}} = 50\kappa$ in a largely molten region with $\phi > 0.4$ (e.g., Ogawa, 2020). $\kappa_{\mathrm{edd}}$ is assumed to gradually increase with increasing $\phi$ as $\phi^3$ (Ogawa, 2018a). We confirmed that doubling $\kappa_{\mathrm{edd}}$ to $100\kappa$ does not strongly affect the histories of radius change and magma flux (Case eddy100; see Supporting Information). We also assumed that the crustal thermal diffusivity is 0.48 times the mantle thermal diffusivity, taking into account the blanket effect of the crust and regolith layers (Ziethe et al., 2009). Magma migration is calculated in the first term of the right-hand side of Eq. (13) as a transport of the latent heat of melting by migrating magma (Katz, 2008, 2022). For more about the energy equation, see the Appendix of U et al. (2026a).

*Table 1. The meanings of the symbols and their values.*

| Symbol | Meaning | Value |
|---|---|---|
| $T_{\mathrm{sur}}$ | Surface temperature | 270 K |
| $T_0$ | Solidus at the surface | 1360 K |
| $r_{\mathrm{p}}$ | Radius of the surface | 1735 km |
| $r_{\mathrm{c}}$ | Radius of the core-mantle boundary (CMB) | 385 km |
| $r_{\mathrm{crst}}$ | Radius of the Moho | 1700 km |
| $\rho_0$ | Reference density | 3300 kg $\mathrm{m}^{-3}$ |
| $\kappa$ | Thermal diffusivity of the mantle | $6 \times 10^{-7}\ \mathrm{m}^2\ \mathrm{s}^{-1}$ |
| $\kappa_{\mathrm{edd}}$ | Eddy diffusivity in largely molten regions | $50\kappa$ at $\phi > 0.4$ |
| $g_0$ | Gravitational acceleration at the surface | 1.6 m $\mathrm{s}^{-2}$ |
| $g_{\mathrm{r}}$ | Gravitational acceleration at the CMB | 0.55 m $\mathrm{s}^{-2}$ |
| $\Lambda_0$ | Latent heat of melting | 657 kJ $\mathrm{kg}^{-1}$ |
| $C_p$ | Specific heat capacity | 1240 J $\mathrm{K}^{-1}\ \mathrm{kg}^{-1}$ |
| $\eta_0$ | Reference viscosity | $1.6 \times 10^{20} - 10^{22}$ Pa s |
| $\eta_{\mathrm{melt}}$ | Melt viscosity | $1 - 20$ Pa s |
| $E_T$ | Sensitivity of viscosity to temperature | $11.3 \times 10^{-3}\ \mathrm{K}^{-1}$ |
| $\phi_0$ | Reference melt-content | 0.05 |

| Symbol | Meaning | Value |
|---|---|---|
| $k_{\phi_0}$ | Reference permeability | $8.6 \times 10^{-15}\ \mathrm{m}^2$ |
| $\alpha_\mathrm{m}$ | Thermal expansivity in the mantle | $3 \times 10^{-5}\ \mathrm{K}^{-1}$ |
| $\alpha_\mathrm{c}$ | Thermal expansivity in the core | $9 \times 10^{-5}\ \mathrm{K}^{-1}$ |

The core is regarded as a heat bath of a uniform temperature $T_\mathrm{c}$ that changes with time as:

$$c_{pc}\rho_\mathrm{core}V_\mathrm{core}\frac{dT_\mathrm{c}}{dt} = -S_\mathrm{CMB}f\,, \tag{15}$$

where $c_{pc} = 675\ \mathrm{J\ K^{-1}\ kg^{-1}}$ (Ziethe et al., 2009) is the specific heat capacity of the core, $\rho_\mathrm{core} = 6200\ \mathrm{kg\ m^{-3}}$ (Kronrod et al., 2022) the core density, $V_\mathrm{core}$ the volume of the core, and $S_\mathrm{CMB}$ the surface area of the core-mantle boundary (CMB). The heat flux at the CMB $f$ is calculated from

$$f = -\frac{1}{S_\mathrm{CMB}}\int_{S_\mathrm{CMB}}\left(k\frac{\partial T}{\partial r}\right)_{r=r_\mathrm{c}} dS\,. \tag{16}$$

In this study, we neglect the internal heating in the core. The initial temperature of the core is $1875\ \mathrm{K}$.

The internal heating rate $q$ in Eq. (13) changes with time as:

$$q = q^*_\mathrm{tr}q_0\exp\left(-\frac{t}{\tau}\right)\,, \tag{17}$$

where $q_0 = 14.7\ \mathrm{pW\,kg^{-1}}$ is the average initial heating rate at 4.4 Gyr ago estimated earlier by a geochemical method (McDonough & Sun, 1995; Lodders, 2003; Taylor & Wieczorek, 2014); $\tau = 1.5$ Gyr is the decay time of HPEs, approximated as an average of the decay times of $^{235}\mathrm{U}$ and $^{40}\mathrm{K}$ (e.g., Kameyama et al., 1996; Ogawa, 2020). The non-dimensional value $q^*_\mathrm{tr}$, which satisfies $\frac{1}{V}\int_V q^*_\mathrm{tr}\ dV = 1$ ($V$ is the volume of the entire calculated region), expresses the effect of transport of HPEs by convecting matrix and migrating magma, and changes with time as

$$\frac{\partial q^*_\mathrm{tr}}{\partial t} + \nabla\cdot(q^*_\mathrm{tr}\mathbf{U}) = -\nabla\cdot[q^*_\mathrm{l}\phi(\mathbf{u}-\mathbf{U})] + \nabla\cdot(\kappa_\mathrm{edd}\nabla q^*_\mathrm{tr})\,. \tag{18}$$

Here, $q^*_\mathrm{l}$ is the internal heating rate in the melt, calculated as

$$q^*_\mathrm{l} = \frac{q^*_{tr}}{D_q+(1-D_q)\phi}\,, \tag{19}$$

where $D_q = 0.01$ is the partition coefficient of HPEs, defined as the ratio of their concentration in the solid phase to that in the melt phase. We added the diffusion term on the right-hand side of Eq. (18) as we did in the energy equation (13).

The bulk composition $\xi_{\mathrm{b}}$ is calculated from the composition of the solid phase $\xi_{\mathrm{s}}$ and that of the liquid phase $\xi_{\mathrm{l}}$ as

$$\xi_{\mathrm{b}} = (1-\phi)\xi_{\mathrm{s}} + \phi\xi_{\mathrm{l}} \tag{20}$$

and changes with time owing to the mass transport by melt and matrix as

$$\frac{\partial \xi_{\mathrm{b}}}{\partial t} + \nabla \cdot (\xi_{\mathrm{b}}\mathbf{U}) = -\nabla \cdot [\xi_{\mathrm{l}}\phi(\mathbf{u}-\mathbf{U})] + \nabla \cdot [\kappa_{\mathrm{edd}}\nabla(\xi_{\mathrm{b}})]\,. \tag{21}$$

The formulations above are similar to those used in a lunar magma ocean model (Y. Zhang et al., 2025), but their model considers the preferential migration of buoyant anorthite-rich melt, leaving behind a compositionally denser, olivine-rich residue. Our model, in contrast, considers the migration of basaltic melt enriched in dense IBC components during post-magma-ocean mantle evolution; the residue left by basaltic melt is less dense than the original material. Because of this difference, the evolution of the mantle considered here is substantially different from that of magma ocean calculated earlier.

A finite-difference numerical code calculates the basic equations under the Boussinesq approximation on a mesh with 64 (radial) × 64 (latitudinal) × 128 (longitudinal) mesh points. The numerical code is based on a smoothing algorithm called "ACuTE" developed in previous works (Kameyama, 2005; Kameyama et al., 2005, 2008); this algorithm is an iterative solution method that combines the multigrid method with the pseudo-compressibility method with local time stepping to find the velocity and pressure fields that simultaneously satisfy the conservation laws of momentum and mass.

## 2.2 The Initial Condition

We assumed the initial condition illustrated in Figure 1 based on earlier numerical studies of the magma ocean and the subsequent mantle overturn. Figure 1a illustrates the initial distribution of the enthalpy $h$ expressed in terms of $T_{eq} \equiv h/C_p$. $T_{eq}$ is approximately equal to the potential temperature $T_{\mathrm{p}}$ listed in Table 2 in most of the mantle. The initial bulk composition in the mantle $\xi_{\mathrm{b}}^{init}$ is assumed to decrease with depth to mimic the accumulation of material enriched in the IBC component in the deep mantle suggested in earlier models of mantle overturn (e.g., Snyder et al., 1992; Elkins-Tanton et al., 2011) as

$$\xi_{\mathrm{b}}^{\mathrm{init}} = 1 - \xi_0 \exp\left[-\frac{r-r_{\mathrm{c}}}{l}\right], \tag{22}$$

where $\xi_0$ is a constant and $l$ represents the thickness of the overturned IBC-rich layer, which is assumed to be $l^*(= l/L) = 1/5.5$ in the reference case based on some earlier overturn models (e.g., H. Li et al., 2019; Zhao et al., 2019). In the crust, $\xi_{\mathrm{b}} = \xi_{\mathrm{crst}} = 1$ is assumed for simplicity. In the calculations presented below, $\xi_{\mathrm{b}}^{\mathrm{init}}$ is always higher than the eutectic composition $\xi_{\mathrm{e}}$.

The initial distribution of the internal heating rate $q_{\mathrm{ini}}$ is given separately for the mantle and the crust. In the mantle, $q_{\mathrm{ini}}$ is calculated from $\xi_{\mathrm{b}}^{\mathrm{init}}$ by a linear interpolation between the olivine-rich material ($\xi_{\mathrm{b}} = 1$) and the IBC-rich material of the eutectic composition ($\xi_{\mathrm{b}} = \xi_{\mathrm{e}} = 0.1$) as

$$q_{\mathrm{ini}} = q_{\mathrm{e}} + (q_{\mathrm{ol}} - q_{\mathrm{e}})\frac{\xi_{\mathrm{b}}^{\mathrm{init}} - \xi_{\mathrm{e}}}{1 - \xi_{\mathrm{e}}}\,, \tag{23}$$

where $q_{\mathrm{ol}} = 2.77\ \mathrm{pW\,kg^{-1}}$ (Yu et al., 2019) and $q_{\mathrm{e}} = 110\ \mathrm{pW\,kg^{-1}}$ are the heating rates of the olivine-rich and IBC-rich materials, respectively. The value of $q_{\mathrm{e}}$ is based on the earlier estimate that the overturned material is 7.5 times more enriched in HPEs than the bulk silicate Moon (Hess & Parmentier, 1995). In the crust, $q_{\mathrm{ini}}$ is uniform, and its value is chosen so that the total amount of HPEs in the crust $Q_{\mathrm{crst}}$ is $F_{\mathrm{crst}}$ times that in the mantle $Q_{\mathrm{mantle}}$. Here, $F_{\mathrm{crst}}$ is related to the concentration ratio $F_{\mathrm{crst}}^{*}$ in Table 2 by $F_{\mathrm{crst}} = F_{\mathrm{crst}}^{*} m_{\mathrm{crst}}/m_{\mathrm{mantle}}$, where $m_{\mathrm{crst}}$ and $m_{\mathrm{mantle}}$ are the masses of the crust and the mantle. The constant $\xi_0$ in Eq. (22) is determined from the total budget of HPEs at 4.4 Gyr ago: $Q_{\mathrm{mantle}} + Q_{\mathrm{crst}} = Q_0$, where $Q_0 = \int_V q_0\ dV \left(= \int_V q_{\mathrm{ini}}\ dV\right)$ is the total heating rate of the silicate Moon calculated from the average heating rate $q_0$ in Eq. (17): here, $V$ is the volume of the entire calculated region. From this budget and Eq. (23), we obtain

$$\frac{Q_0}{1 + F_{\mathrm{crst}}} = Q_{\mathrm{mantle}} = \int_{V_m} q_{\mathrm{ini}}\, dV, \tag{24}$$

which we calculated for $\xi_0$ with assumed values of $F_{\mathrm{crst}}^{*}$ and $l^{*}$; here $V_m$ is the volume of the mantle part of the calculated region.

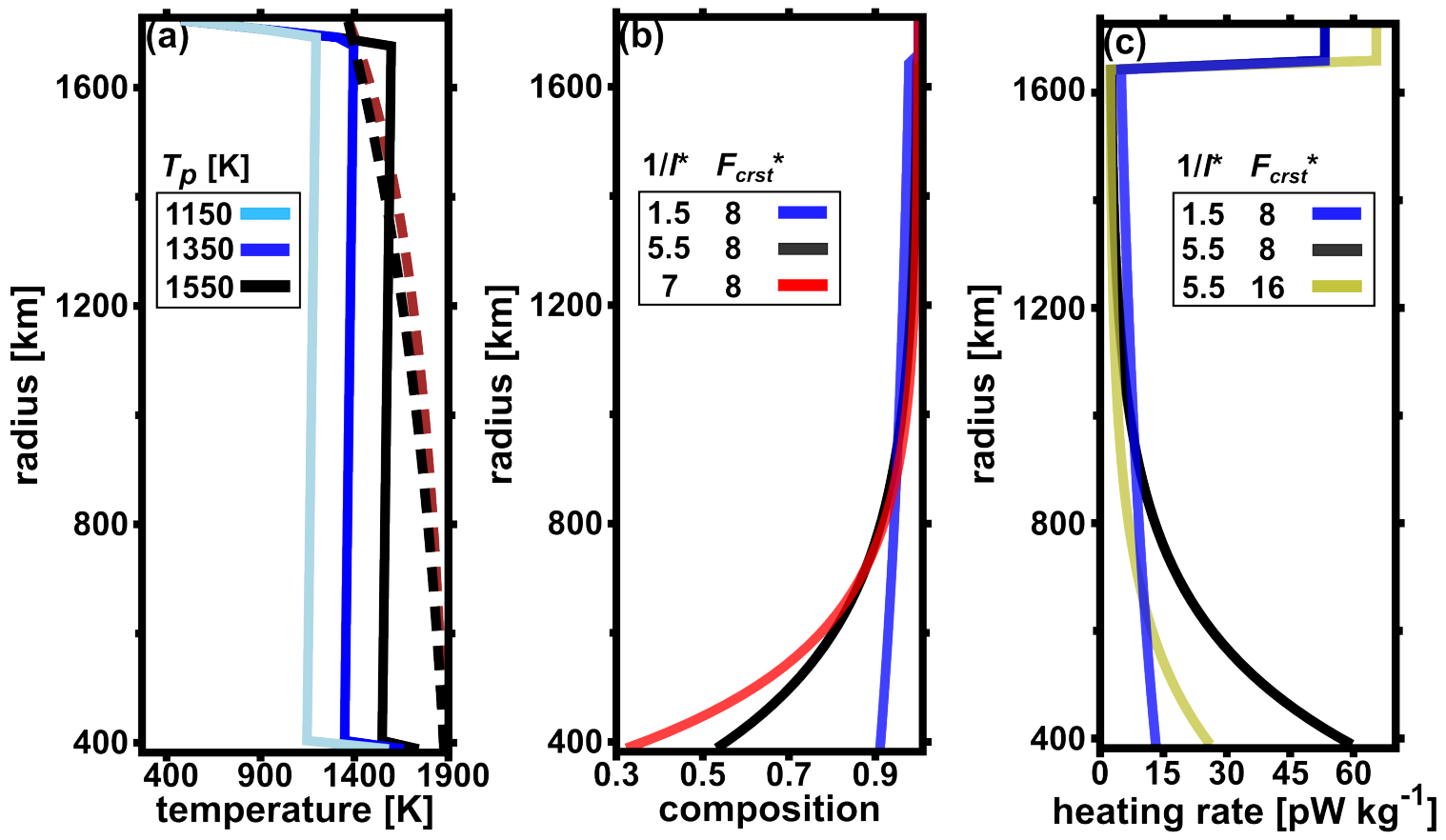

Figure 1. An illustration of the initial distributions of (a) potential temperature $h/C_p$, (b) bulk composition $\xi_{\mathrm{b}}$, and (c) internal heating rate $q$. In (a), the black and brown dashed lines are the solidus for all cases except Case inv-vl and that for Case inv-vl, respectively (see Section 4.2.3). The meanings of the variable parameters $T_{\mathrm{p}}$, $l^{*}$, and $F_{\mathrm{crst}}^{*}$ are described in Table 2.

*Table 2. The variable parameters of the numerical models.*

| Parameter | Meaning | Range of values |
|---|---|---|
| $M$ | Dimensionless reference permeability | $5$ to $100$ |
| $Ra$ | Rayleigh number | $2.15 \times 10^4$ to $10^6$ |
| $E_T^*$ | Dimensionless sensitivity of viscosity to temperature | $3$ to $9$ |
| $T_\mathrm{p}$ | Initial potential temperature in the mid-mantle | 1150 to 1750 K |
| $l^*$ | Thickness of the overturned layer after mantle overturn | 1/7 to 1/1.5 |
| $F_\mathrm{crst}^*$ | Concentration ratio of HPEs in the crust to that in the mantle | 8 to 16 |

### 2.3 The Parameter Values

We carried out numerical experiments at various values of the non-dimensional reference permeability $M \equiv \frac{k_{\phi_0}\rho_0 g_0 L}{\kappa\eta_\mathrm{melt}}$ (see Eq. (A3)) and Rayleigh number $Ra \equiv \frac{\rho_0\alpha_\mathrm{m}(\Lambda_0/C_p)g_0 L^3}{\kappa\eta_0}$ (see Eq. (A6)), as well as the sensitivity of viscosity to temperature $E_T^* = E_T\Lambda_0/C_p$, the initial potential temperature of the mantle $T_\mathrm{p}$ (see Figure 1a), the thickness of the overturned layer $l^* = l/L$, and the ratio of the concentration of HPEs in the crust to that in the mantle $F_\mathrm{crst}^*$. The non-dimensional reference permeability $M$ is proportional to $k_{\phi_0}/\eta_\mathrm{melt}$ and expresses how easily magma migrates upward. We varied the values of $M$ in the range of $5 \le M \le 100$, which corresponds to the range of the viscosity $\eta_\mathrm{melt}$ of 1 to 20 Pa s (Miller et al., 2014). On the other hand, the Rayleigh number $Ra$ is inversely proportional to the reference viscosity and expresses how readily mantle convection occurs. We estimated that this value is in the range of $Ra$ = $2.15 \times 10^4$ to $10^6$ corresponding to $\eta_0$ = $1.6 \times 10^{22}$ to $10^{20}$ Pa s, based on an earlier estimate of mantle rheology (Karato & Wu, 1993). This range of $Ra$ is below the threshold for the onset of thermal convection driven by basal heating (Yanagisawa et al., 2016).

Other parameters listed in Table 2 were also varied to see how the numerical results depend on the thermal and compositional structure of the mantle assumed in the initial conditions. The range of $T_\mathrm{p}$ is taken from that of initial temperature in Laneuville et al. (2013) and that of post-overturn temperature in some models of the mantle overturn (Boukaré et al., 2018; H. Li et al., 2019; Yu et al., 2019). $l^*$ is estimated from the density of the IBC-enriched basal layer calculated in some earlier numerical models of mantle overturn (e.g., H. Li et al., 2019; Zhao et al., 2019). The overturned layer is thinner and more enriched in HPEs and the IBC component at a lower $l^*$. We also took the value of $F_\mathrm{crst}^*$ from the estimates in Konrad & Spohn (1997) and Spohn et al. (2001).

## 3 Two Modes of Magma Migration

The earlier linear perturbation analysis and numerical studies of the basic equations in Section 2.1 (Eqs. (7), (8), (10), (11), and (13)) carried out in U et al. (2026a) predict that magma migrates in a partially molten mantle in two modes depending on the retention number $R_t$ defined by

$$R_t = \frac{\eta_{\mathrm{melt}} L^2}{\eta_0 k_{\phi 0}} = \frac{C_p}{\alpha_m \Lambda_0} \frac{Ra}{M}. \qquad (25)$$

(We summarized the linear perturbation analysis in Appendix B.) $R_t$ expresses how important the convective flow of matrix is relative to the percolative flow of magma. For the ranges of $\eta_{\mathrm{melt}}$ and $\eta_0$ estimated for the lunar mantle (Section 2.3), $R_t$ ranges from about $10^4$ to $3 \times 10^7$.

At a high $R_t$, a positive feedback called the magmatism-mantle upwelling feedback through melt buoyancy (MMUb feedback; Ogawa, 2018b, 2020) or the buoyant melting instability (e.g., Tackley & Stevenson, 1993; Hernlund et al., 2008a, b; Pusok et al., 2022) dominates the flow in the mantle that contains partially molten regions (Figure 2a): the buoyancy of melt generated in an upwelling region of the mantle by decompression melting further accelerates the upwelling (Figure 2a-1). In Figure 2a-2 where a partially molten layer lies beneath a solid layer, for example, melt and matrix in the partially molten plumes ascend together through the solid layer while generating more melt in the plumes. This feedback becomes substantial when

$$R_t \gg R_{tb} \equiv \frac{6\bar{\phi}^2}{\phi_0^3} \left(\frac{2\pi L}{d}\right)^2, \qquad (26)$$

where $\bar{\phi}$ is a typical melt content of a partially molten region and $d$ is its typical size (see Appendix B; see also Figure 2 and Eq. (43) in U et al., 2026a).

At $R_t \ll R_{tb}$, in contrast, melt percolates upward through the matrix, and the melt percolation-dominated instability (also called solitary waves; Richter & Daly, 1989) dominates the flow in a partially molten mantle (Figure 2b). A small perturbation in the melt content $\phi$ makes the upward flux of percolating melt vary from place to place. The matrix expands where the incoming melt accumulates and contracts where the melt leaves (Eqs. (7), (10), and (11)), allowing the perturbation to move upward and grow (Figure 2b-1). When a partially molten layer lies beneath a solid layer, melt accumulates along the top of the partially molten layer (Figure 2b-2), and the instability makes the melt-enriched layer undulate and break up into isolated patches of melt, called magma fingers, that rise into the solid layer. Within the magma fingers, the matrix moves downward while the melt moves upward.

When $R_t \approx R_{tb}$, both the MMUb feedback and the melt percolation-dominated instability are important and interact with each other. Magma fingers first grow from the top of a partially molten layer. As the fingers grow, the flow of matrix within them gradually turns from downward to upward, and the fingers eventually evolve into partially molten plumes (Figure 2c; U et al., 2026a).

For $\bar{\phi} = 0.1$ and $d/L = 0.2$, which are typical values for the Moon, $R_{tb}$ is about $5 \times 10^5$, which is within the range of $R_t$ estimated above for the Moon. Based on these estimates of $R_t$ and $R_{tb}$, we calculated most of our cases, including Case Ref, at $R_t$ around $10^6$, which is close to $R_{tb}$. To see how the mode of magma migration affects the mantle evolution, we also calculated two cases in which $R_t$ is far from $R_{tb}$ (Table 3): in Case M5, $R_t$ is about $3 \times 10^7$, much higher than $R_{tb}$; in Case Ra2e4, $R_t$ is about $10^4$, much lower than $R_{tb}$ (Section 4.2).

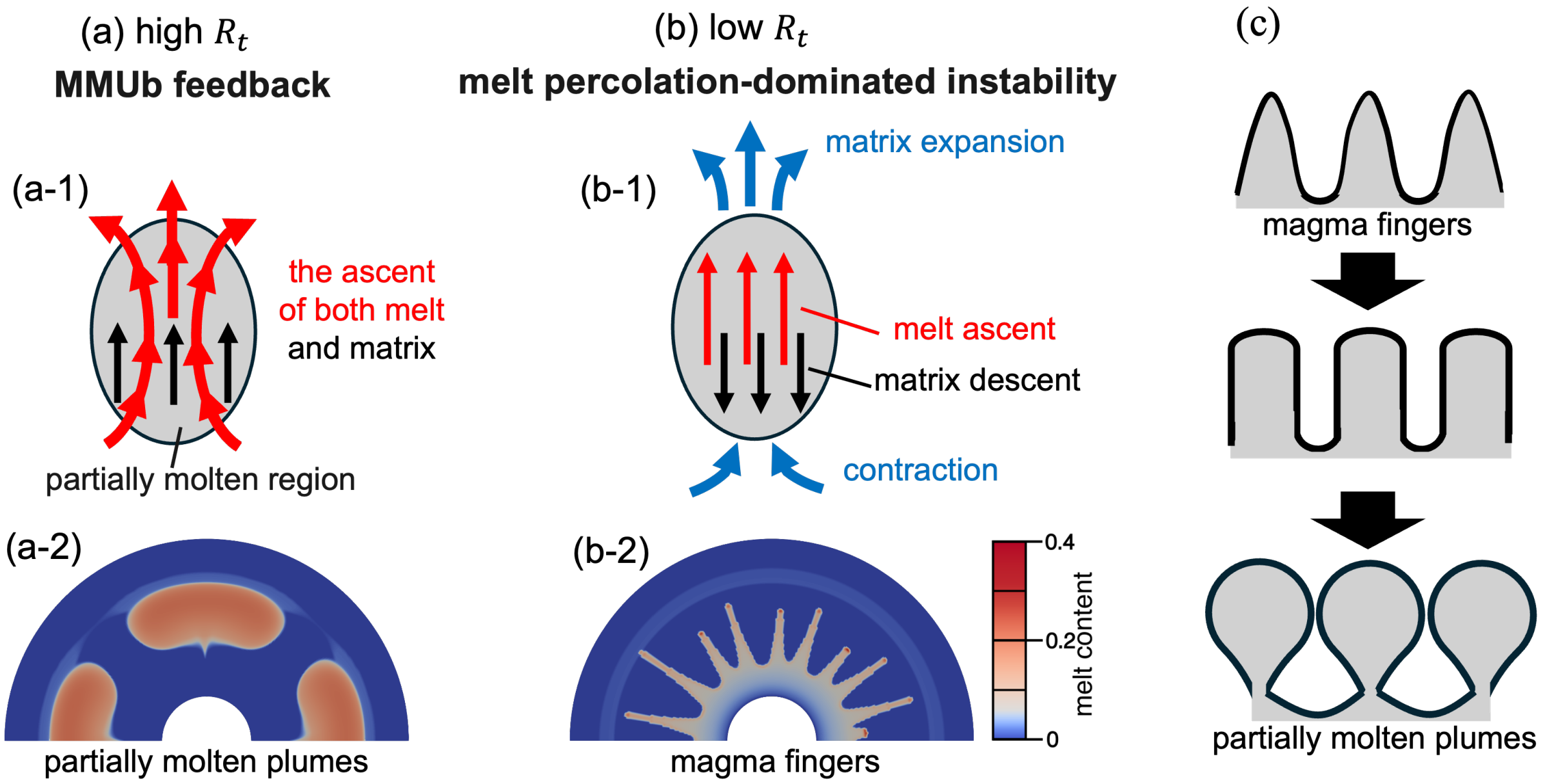

Figure 2. (a-1, b-1) Schematic illustrations of two modes of magma migration that arise in a partially molten mantle depending on the retention number $R_t$: MMUb feedback at a high $R_t$ and melt percolation-dominated instability at a low $R_t$. Also shown are snapshots of melt-content distributions that arise when (a-2) partially molten plumes due to the MMUb feedback and (b-2) magma fingers due to the percolation-dominated instability occur. These snapshots are taken from U et al. (2023). (c) Schematic illustration of transition from magma fingers to partially molten plumes.

# 4 Results

## 4.1 The Reference Case

Figures 3-6 as well as Movies S1, S2 in the Supporting Information show the reference case (Case Ref) calculated at the reference permeability $M = 100$, corresponding to $\eta_{\mathrm{melt}} = 1$ Pa s; the Rayleigh number $Ra = 2.15 \times 10^6$, corresponding to $\eta_0 = 1.6 \times 10^{20}$ Pa s; the sensitivity of viscosity to temperature $E_T^* = 6$; the initial potential temperature of the mantle $T_{\mathrm{p}} = 1550$ K; the initial thickness of the overturned layer $l^* = 1/5.5$; the initial ratio of HPE concentration in the crust to that in the mantle $F_{\mathrm{crst}}^* = 8$ (Table 2). The corresponding retention number $R_t$, about $10^6$, is close to the threshold $R_{tb}$ in Eq. (26) as described above; both magma fingers and partially molten plumes are expected to contribute to the mantle evolution and the resulting history of magmatism.

Table 3. The values of the parameters listed in Table 2 and a summary of results.

| Case # | $M$ | $Ra$ | $E_T^*$ | $T_{\mathrm{p}}$ [K] | $1/l^*$ | $F_{\mathrm{crst}}^*$ | $\Delta R_{peak}$ [km] | $t_{\Delta R}$ [Gyr] | $\chi_{Gyr}$ [km Gyr$^{-1}$] | $L_{30}$ [km] | $L_{60}$ [km] |
|---|---|---|---|---|---|---|---|---|---|---|---|
| Ref | 100 | $2.15 \times 10^6$ | 6 | 1550 | 5.5 | 8 | 0.80 | 0.74 | -0.87 | 4.99 | 8.81 |
| M5 | 5 | $2.15 \times 10^6$ | 6 | 1550 | 5.5 | 8 | 1.83 | 1.17 | -0.80 | 0.00 | 0.46 |
| M50 | 50 | $2.15 \times 10^6$ | 6 | 1550 | 5.5 | 8 | 1.17 | 0.80 | -0.86 | 1.14 | 4.74 |
| Ra2e4 | 100 | $2.15 \times 10^4$ | 6 | 1550 | 5.5 | 8 | 0.93 | 1.20 | -0.88 | 0.48 | 1.30 |
| inv-vl | 100 | $2.15 \times 10^6$ | 6 | 1550 | 5.5 | 8 | 1.74 | 1.20 | -0.79 | 0.00 | 7.66 |
| melt-dep | 100 | $2.15 \times 10^6$ | 6 | 1550 | 5.5 | 8 | 0.57 | 0.68 | -0.85 | 9.75 | 13.91 |
| IBC-dep | 100 | $2.15 \times 10^6$ | 6 | 1550 | 5.5 | 8 | 0.72 | 0.73 | -0.53 | 3.78 | 8.33 |
| eddy100 | 100 | $2.15 \times 10^6$ | 6 | 1550 | 5.5 | 8 | 0.72 | 0.72 | -0.83 | 7.05 | 9.66 |
| T1150 | 100 | $2.15 \times 10^6$ | 6 | 1150 | 5.5 | 8 | 1.23 | 2.76 | -0.07 | — | — |
| T1350 | 100 | $2.15 \times 10^6$ | 6 | 1350 | 5.5 | 8 | 0.64 | 1.75 | -0.49 | 0.05 | 0.32 |
| T1750 | 100 | $2.15 \times 10^6$ | 6 | 1750 | 5.5 | 8 | 1.17 | 0.34 | -1.14 | 5.04 | 14.72 |
| F16 | 100 | $2.15 \times 10^6$ | 6 | 1550 | 5.5 | 16 | 0.16 | 1.34 | -0.82 | 0.07 | 0.10 |
| L1.5 | 100 | $2.15 \times 10^6$ | 6 | 1550 | 1.5 | 8 | 0.32 | 1.40 | -0.82 | 1.00 | 4.31 |
| L3.5 | 100 | $2.15 \times 10^6$ | 6 | 1550 | 3.5 | 8 | 1.08 | 0.96 | -0.81 | 0.49 | 1.49 |
| L7 | 100 | $2.15 \times 10^6$ | 6 | 1550 | 7 | 8 | 0.50 | 0.74 | -0.84 | 5.94 | 9.98 |
| Et3 | 100 | $2.15 \times 10^6$ | 3 | 1550 | 5.5 | 8 | 0.79 | 0.88 | -0.98 | 0.95 | 6.78 |
| Et9 | 100 | $2.15 \times 10^6$ | 9 | 1550 | 5.5 | 8 | 0.90 | 0.75 | -0.84 | 6.14 | 9.38 |
| Ref-grad | 100 | $2.15 \times 10^6$ | 6 | 1550 | 5.5 | 8 | 0.78 | 0.83 | -0.84 | 1.90 | 5.41 |
| M5-grad | 5 | $2.15 \times 10^6$ | 6 | 1550 | 5.5 | 8 | 1.75 | 1.13 | -0.79 | 0.00 | 0.06 |
| Ra2e4-grad | 100 | $2.15 \times 10^4$ | 6 | 1550 | 5.5 | 8 | 0.94 | 1.19 | -0.89 | 0.57 | 1.39 |

Note. $\Delta R_{peak}$ and $t_{\Delta R}$ stand for the magnitude and timing of the peak of radial expansion, respectively, where $\Delta R_{peak}$ is measured from the minimum of $\Delta R$ that precedes the expansion; $\chi_{Gyr}$ the contraction rate for the past 1 Gyr; $L_{30}$ and $L_{60}$ are the global equivalent thicknesses of the total amount of magma that crossed upward the depth levels of 30 km and 60 km for the calculated 4.4 Gyr period. The suffix -grad stands for the cases where $g_0$ in Eq. (8) is replaced by the depth-dependent $g$ (see Text S1 in the Supporting Information).

#### 4.1.1 Thermal and Structural Evolution of the Mantle

For the first 300 Myr, the temperature calculated in Case Ref rises at the base of the mantle, which is initially enriched in HPEs, and magma is generated there (Figure 3a and the red line in Figure 4a). At 0.4 Gyr, the distributions of melt content, bulk composition, and temperature are still laterally uniform (Figure 3). However, finger-like structures, or magma fingers, eventually develop along the top boundary of the partially molten layer and grow upward to transport the IBC component after 0.4 Gyr (Figure 3a to c for 0.56 Gyr; see also Movies S1, S2). These magma fingers grow by the melt percolation-dominated instability described in Section 3 (Figure 2b). As expected for this instability, the matrix flows downward within the heads of the magma fingers (see the white arrows in Figure 5 for 0.56 Gyr), while it flows upward above the heads (see the yellow arrows). Partially molten plumes then develop along the fingers and ascend to the uppermost mantle (Figure 3a, c for 0.56 and 0.70 Gyr; see also Movies S1, S2). The plumes are driven by the buoyancy of melt. The matrix ascends together with the melt in the plumes (see the light-brown iso-surfaces in Figures 3a and 5 for 0.70 Gyr), and decompression melting in the ascending plumes generates more melt, whose buoyancy further accelerates the ascent; that is, the MMUb feedback described in Section 3 operates in the plumes (Figure 2a, c). The plumes induce extensive partially molten regions in the uppermost mantle (Figure 3a) and a peak in the plot of rms-velocity (Figure 4c). When the plumes reach the depth level of 150 km in the lithosphere, above which the permeability remains non-zero even when $\phi = 0$ (see Eq. (12)), hat-like partially molten heads develop at their tops as the magma percolates upward out of the plumes (see the arrows in Figure 3a for 0.70 Gyr; see also Movie S2). These heads keep the uppermost mantle above the plumes hot and permeable, and the magma supplied by the plumes percolates through the heads to the depth levels of 60 and 30 km.

After 0.70 Gyr, most of the partially molten plumes that have ascended to the uppermost mantle gradually solidify as they are cooled from the surface boundary by thermal diffusion (Figures 3a, c for 0.70 and 1.99 Gyr). The IBC material transported to the uppermost mantle forms basaltic blobs upon solidification (referred to as IBC blobs hereinafter; see the arrows in Figures 3b and 5 for 0.70 Gyr). Deeper parts of the compositionally dense IBC blobs then founder to the CMB owing to the Rayleigh-Taylor instability (see the arrows in Figures 3b and 5 for 1.99 Gyr; see also Movie S1). The sinking IBC blobs cause upward counterflows of matrix, generating magma, not enriched in HPEs and the IBC component, by decompression melting (see the arrows in Figure 3a for 2.36 Gyr and the slight increase in the rms-velocity at around 2 Gyr in Figure 4c). After this plume activity caused by the foundering IBC blobs, the mantle is cooled, partially molten regions shrink, and mantle convection becomes weaker with time (Figures 3a and 4). The horizontally averaged temperature is below the solidus temperature in the entire mantle at 4.4 Gyr (Figure 4d). Some regions at $r$ ~1000 km are, however, still partially molten with low melt content ($\phi < 5\%$) as shown in Figure 3a for $t$ =4.34 Gyr and Figure 4d.

An interesting feature of the calculated convective flow is that the partially molten plumes do not laterally move so much, and that their locations are rather fixed as can be seen from Figures 3, 5, and Movie S1. Once a partially molten plume ascends through the mantle, the subsequent upwelling flows tend to occur through the hotter and softer conduit formed by the earlier plume. Indeed, the locations of elevated temperature shown in Figure 3c as well as those of upward convective flow shown in Figure 5 remain almost the same after 0.70 Gyr.

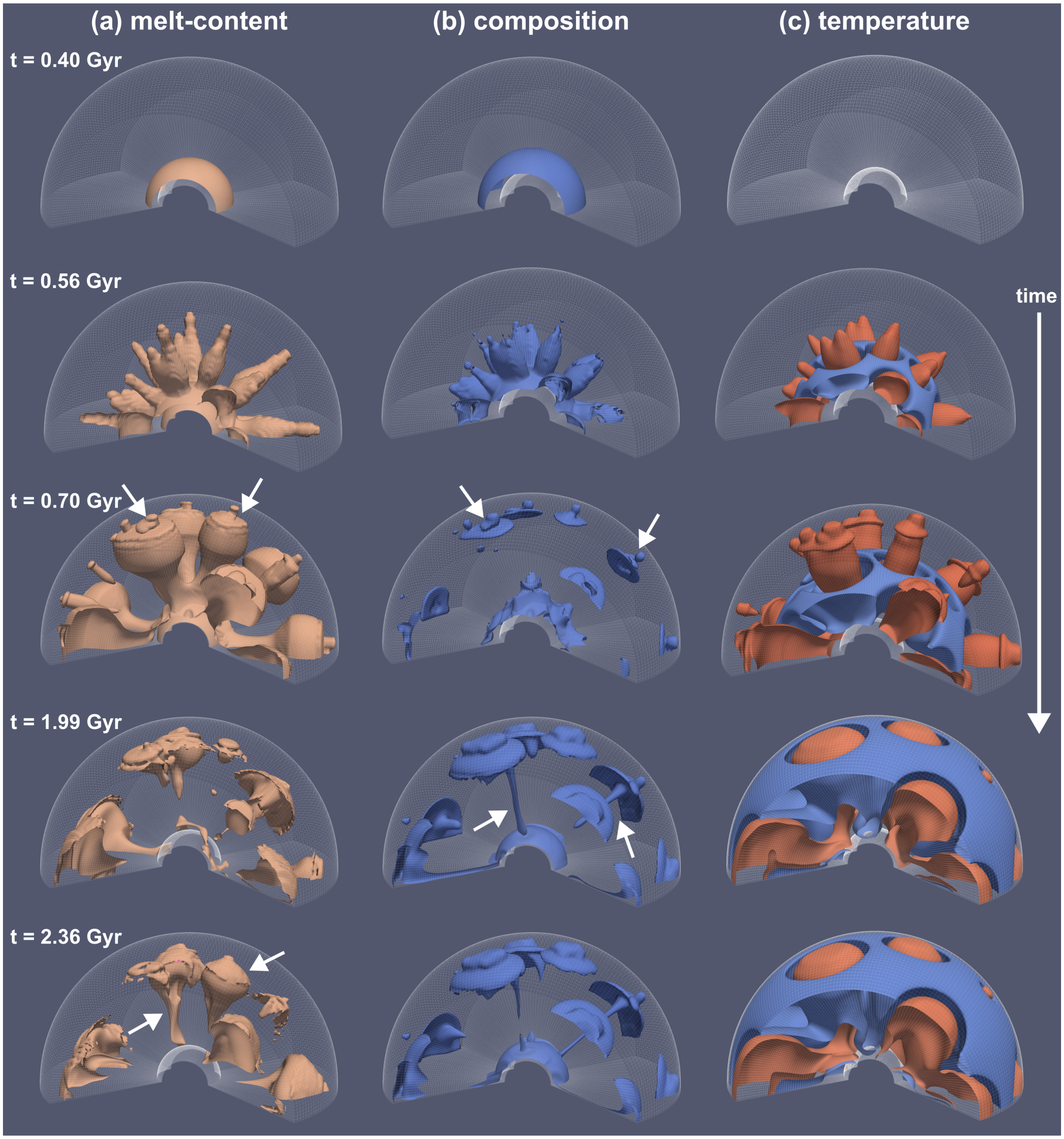

Figure 3. Snapshots of the distributions of (a) melt-content $\phi$ with the iso-surface of $\phi = 1.5\%$, (b) bulk composition $\xi_b$ with the iso-surface of $\xi_\mathrm{b} = 0.85$, and (c) the temperature deviation from the horizontal average $\delta T$ with the iso-surfaces of $\delta T = 20\ K$ (the red surface) and $\delta T = -20$ K (the blue surface) calculated for Case Ref where the permeability $M = 100$; $Ra = 2.15 \times 10^6$, corresponding to $\eta_0 = 1.6 \times 10^{20}$ Pa s; $E_T^* = 6$; $T_\mathrm{p} = 1550$ K; $l^* = 1/5.5$; $F_\mathrm{crst}^* = 8$. The elapsed time is indicated in the figure.

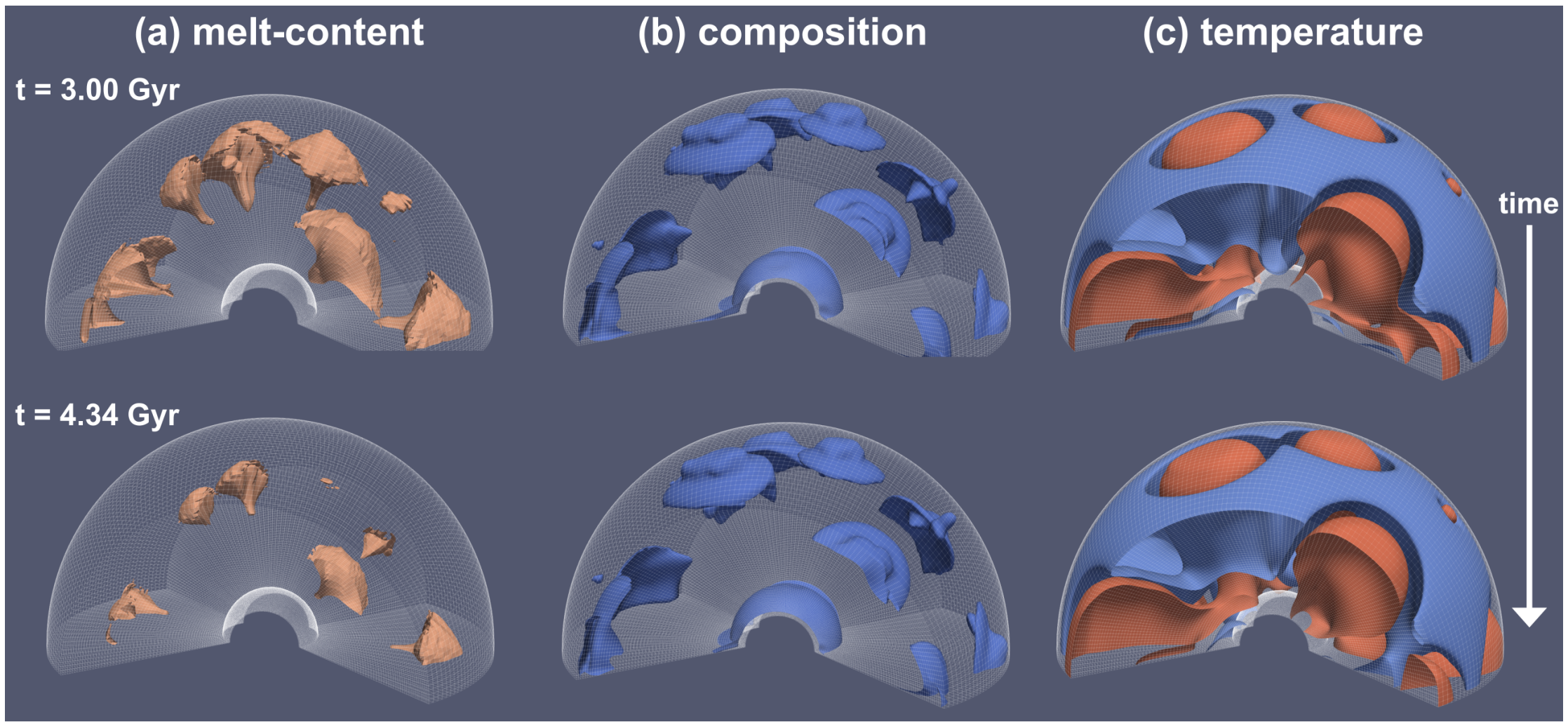

Figure 3. Continued.

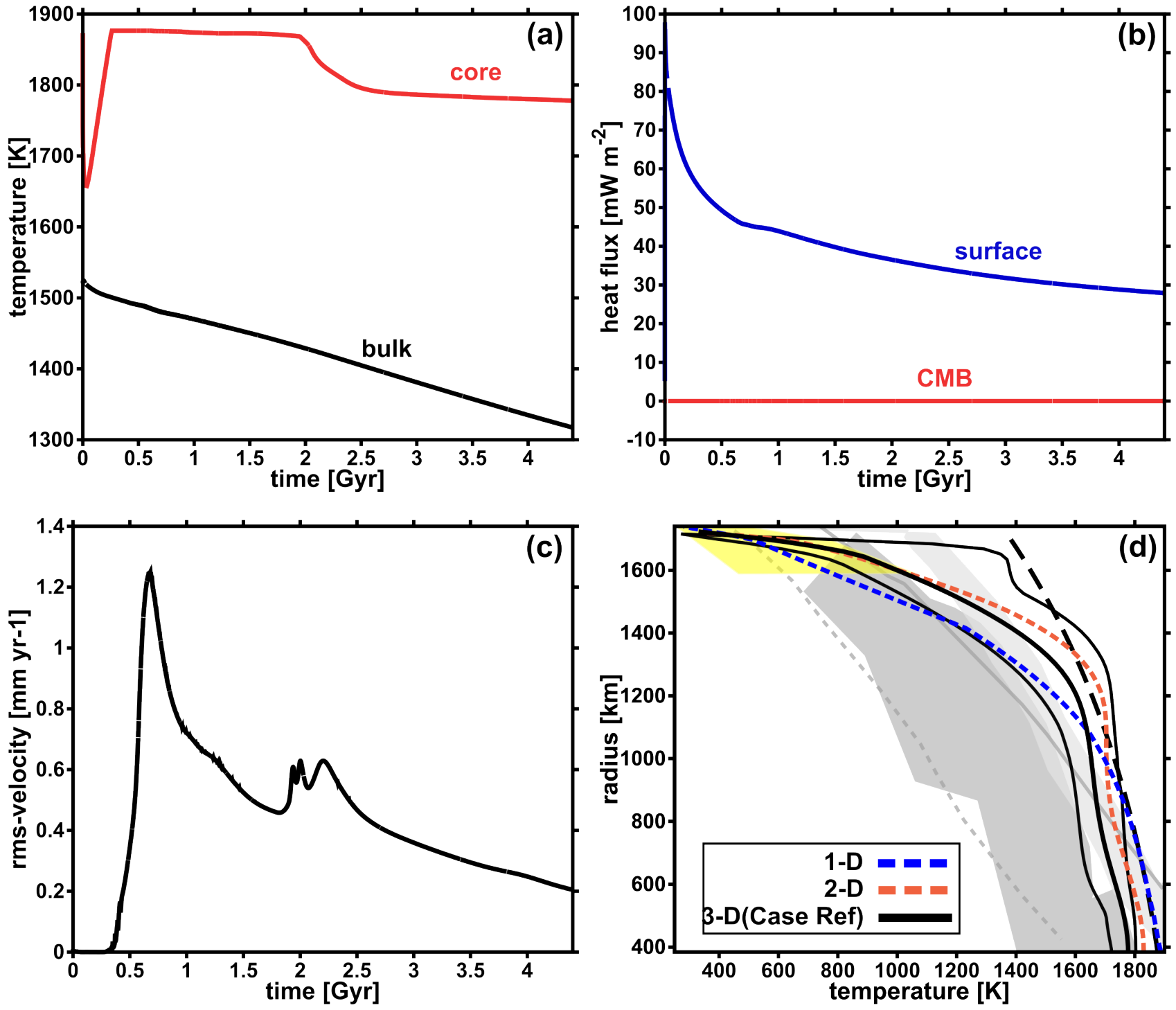

Figure 4. (a) The average temperature in the mantle (bulk) and the temperature of the core, which is equal to the temperature at the core-mantle boundary (CMB); (b) the horizontal average of heat flux on the surface and the CMB; (c) the root-mean-square average of matrix velocity in the mantle, all plotted against time. (d) The horizontal average as well as the maximum and minimum of the temperature at each height calculated at 4.4 Gyr for Case Ref (the thick black line and thin black lines, respectively); the black dashed line is the solidus. Also shown are the horizontally averaged temperatures at 4.4 Gyr for the reference cases of earlier

1-D (the dashed blue line; U et al., 2022) and 2-D models (the dashed red line; U et al., 2023). The gray and light-gray areas are the temperature distributions in today's lunar mantle inferred by Khan et al. (2006) and Khan et al. (2014), respectively; the yellow areas are those estimated from the heat flux at the surface (Siegler & Smrekar, 2014; Siegler et al., 2022); the gray lines are those estimated under the assumption that the mantle consists of dry olivine (the solid line) and wet olivine with 0.01 wt % $H_2O$ (the dashed line) by Karato (2013).

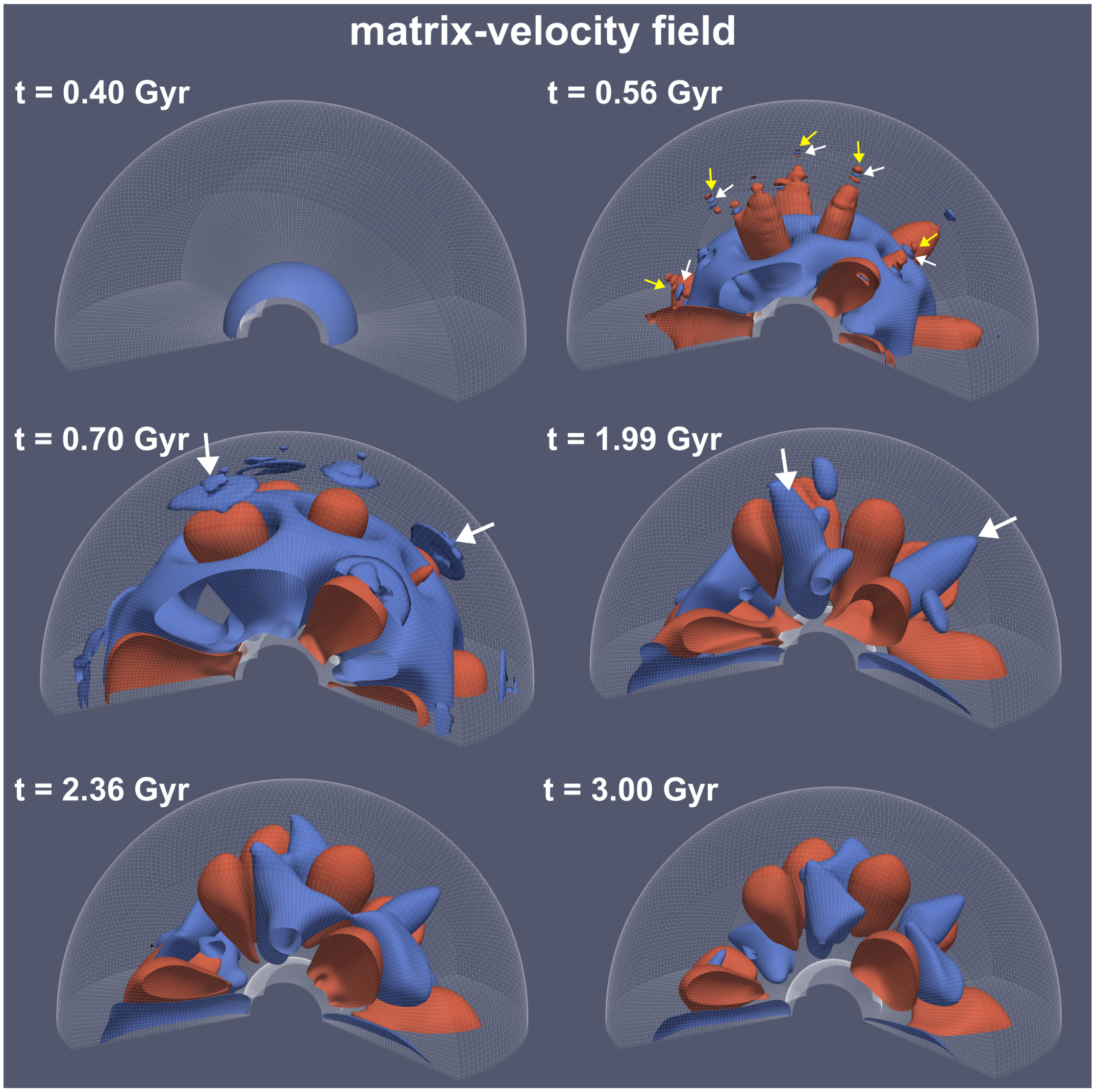

Figure 5. Snapshots of the radial component of matrix-velocity $U_r$ field with the iso-surfaces of $2.8 \times 10^{-1}$ (the red surface) and $-2.8 \times 10^{-1}$ (the blue surface) mm yr$^{-1}$ calculated for Case Ref.

#### 4.1.2 Radius Change

Figure 6a shows how the radius of the Moon modeled here changes with time owing to the mantle evolution shown in Figures 3 to 5. In this estimate of radius change $\Delta R$, we consider the effect of both melting and thermal expansion:

$$\Delta R = \frac{1}{r_\mathrm{p}^2}\left[\frac{\alpha_\mathrm{c}\Delta T_\mathrm{c}}{3} r_\mathrm{c}^3 + \int_{V_\mathrm{m}} \left(\alpha_\mathrm{m}\Delta T + \frac{\Delta V_\mathrm{l}}{V_0}\Delta\phi\right) dV_\mathrm{m}\right]. \quad (27)$$

The first and second terms in the bracket of the right-hand side represent the volume changes of the core and the mantle, respectively. Here, $\alpha_\mathrm{c}$ is the thermal expansivity of the core, $\Delta T_\mathrm{c} \equiv T_\mathrm{c}(t) - T_\mathrm{c}(t=0)$ is the deviation of the temperature $T_\mathrm{c}$ in the core at time $t$ from its initial value $T_\mathrm{c}(t=0)$, and $\Delta T \equiv T(r,\theta,\psi,t) - T(r,\theta,\psi,t=0)$ and $\Delta\phi \equiv \phi(r,\theta,\psi,t) - \phi(r,\theta,\psi,t=0)$ are the deviations in the temperature $T$ and melt content $\phi$ in the mantle from their initial values, respectively. We assume that the initial condition in this model is the state immediately after mantle overturn (see Section 2.2), because the tectonic features of expansion/contraction in the crust are not recorded during the era of the magma ocean (Elkins-Tanton & Bercovici, 2014).

The blue and red lines in Figure 6a show the contributions of thermal expansion/contraction and melting in Eq. (27) to the radius change, respectively. Owing to the thermal effect, the radius contracts monotonically by about 3.5 km as the mantle is cooled throughout the calculated 4.4 Gyr history (the blue line in Figure 6a). On the other hand, the red line, which shows the radius change by melting, indicates that the Moon expands by more than 1 km for the first 0.8 Gyr owing to the widespread partial melting of the mantle caused by magma fingers and partially molten plumes (Figure 6a). Thereafter, the Moon gradually contracts with time as the partially molten regions shrink: the temporary expansion of partially molten plumes caused at 2.36 Gyr (see the arrows in Figure 3a) does not cause any noticeable radius increase. As a whole, the sum of the two effects shown by the black line in Figure 6a indicates that the Moon expands radially by about 0.8 km for the first 0.8 Gyr and then contracts until the end of the calculated history. The contraction rate is around $0.9\ \mathrm{km\ Gyr^{-1}}$ for the past 1 Gyr (see $\Delta R_\mathrm{peak}$, $t_{\Delta\mathrm{R}}$, and $\chi_\mathrm{Gyr}$ in Table 3).

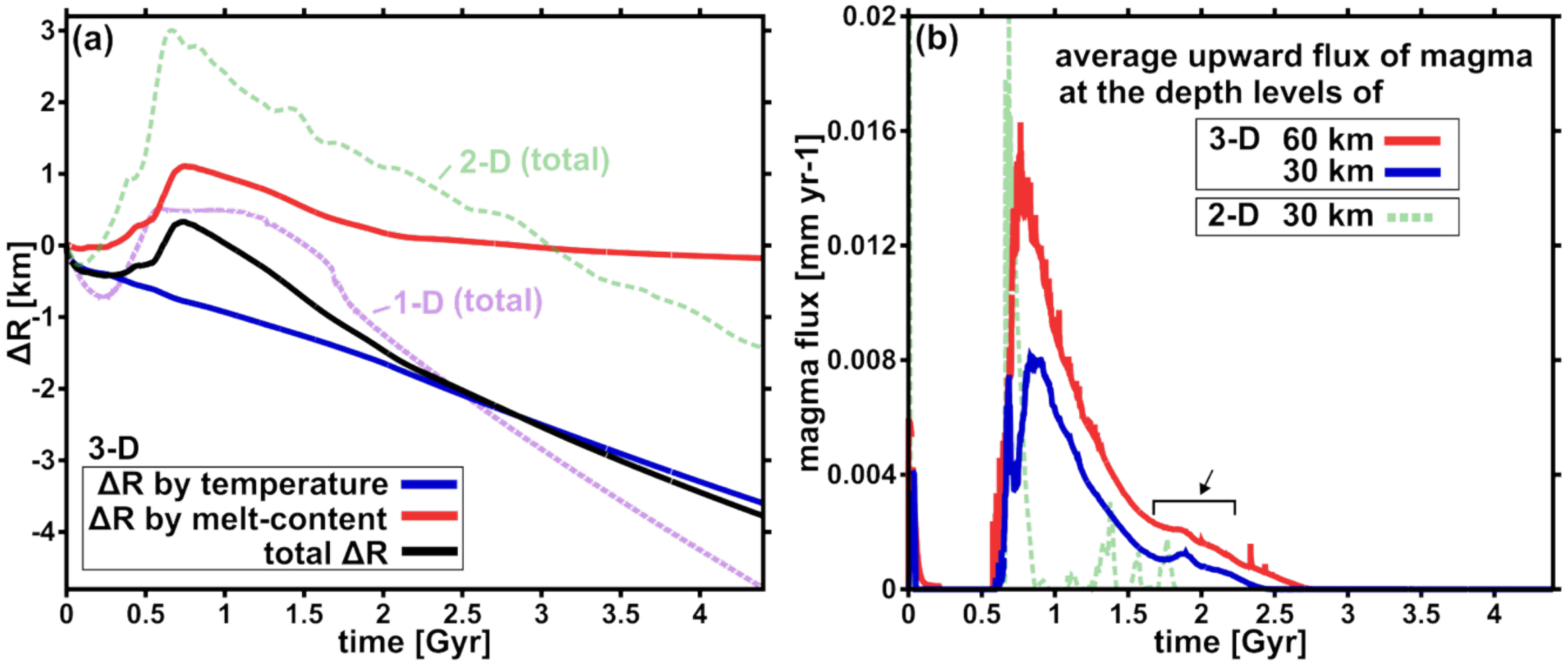

Figure 6. (a) The history of the radius change calculated for the reference case (Case Ref). The radius change $\Delta R$ is defined by Eq. (27); the blue and red lines indicate the contributions of thermal expansion/contraction and melting, respectively, to the total radius change (the black line); the dotted purple and light-green lines show the total radius change in the 1-D model of U et al. (2022) and the 2-D model of U et al. (2023), respectively. (b) The horizontally averaged upward magma flux ($u_r \phi$) at the depth level of 30 km (the blue line) and 60 km (the red line) calculated in the reference case (Case Ref) plotted against time. Also plotted is the result of a 2-D model of U et al. (2023) (the dotted light-green line); the magma flux is zero at the depth level of 30 km and 60 km in the 1-D model of U et al. (2022).

#### 4.1.3 The Flux of Magma

Figure 6b shows the upward flux of magma across the depth level of 30 km (the blue line) and 60 km (the red line) calculated in the reference case; both fluxes follow a similar evolution. The magma flux is significant at first because of the partial melting in the uppermost mantle assumed in the initial condition (Figure 1) and then decreases to 0 mm $\mathrm{yr}^{-1}$ within the first 0.3 Gyr as the uppermost mantle solidifies due to the cooling from the surface boundary. The magma flux, however, becomes substantial with a peak at 0.8 Gyr (Figure 6b) as the partially molten plumes generated in the deep mantle ascend to the uppermost mantle (Figure 3a for 0.70 Gyr). The magma flux then declines but persists until around 2.5 Gyr because of the persistent partially molten plumes observed in Figure 3. The resurgence of magma ascent caused by the descending IBC blobs (Figure 3 for 1.99 Gyr and 2.36 Gyr) contributes to slowing the decline of magma flux at around 2 Gyr (see the arrow in Figure 6b). If all of the magma that crossed the depth level of 30 km erupted onto the surface, the global equivalent thickness of the lava layer would be 5 km ($L_{30}$ in Table 3).

### 4.2 Roles of Partially Molten Plumes and Magma Percolation

To understand more clearly how partially molten plumes and the percolation of magma contribute to the mantle evolution in Case Ref described above, we further calculated Case M5 at a lower permeability (Section 4.2.1), Case Ra2e4 at a lower Rayleigh number (Section 4.2.2), and Case inv-vl where magma is denser than the coexisting matrix in the deep mantle (Section 4.2.3). The parameter values and a summary of these cases are given in Table 3.

#### 4.2.1 The Case Dominated by Partially Molten Plumes (Case M5)

We first recalculated the model at a lower reference permeability of $M = 5$ (Case M5 in Figure 7 and Movie S3). The retention number of this case is about $3 \times 10^7$, 20 times higher than that of Case Ref. At this high retention number, the analysis of U et al. (2026a) predicts that partially molten plumes driven by the MMUb feedback dominate the evolution of the mantle, while melt percolation that induces magma fingers becomes negligible (Section 3).

The snapshots in Figure 7 and Movie S3 indeed show that magma fingers do not develop in this case. A partially molten layer develops along the CMB by 0.5 Gyr owing to the strong internal heating as in Case Ref. Then, partially molten plumes enriched in the IBC component and the HPEs directly rise from the layer because the layer is less dense than the solid mantle above it because of melt buoyancy. The plumes ascend to the base of the lithosphere by 0.7 Gyr, inducing a strong convective flow in the mantle (see Figure 7 and the rms-velocity shown by the red line in Figure 8c). In contrast to Case Ref shown in Figure 3, however, the magma in the plume heads does not segregate from them to further penetrate through the lithosphere to the depth level of 60 km as a permeable flow because of the assumed low permeability (see Figure 7 for 0.75, 1.34 Gyr and the red curve in Figure 8f); the magma is cooled by conduction from the surface and solidifies before it reaches this depth level. Instead, the partially molten plume heads laterally spread. Eventually, a limited amount of magma crosses the depth level of 60 km between around 1.5 and 2.5 Gyr when the partially molten plumes heat the lithosphere and their heads directly ascend to this depth level (Figure 8f). The tops of the plume heads then solidify upon cooling from the surface boundary to form thin IBC blobs, but the main parts of the plume heads remain partially molten in the mid-mantle even after 3 Gyr (Figure 7 for 1.34, 3.40 Gyr). The IBC blobs are thin and stick to the base of the stagnant lithosphere; they do not founder, unlike in Case Ref (see Figure 7b for 3.40 Gyr). As a consequence, the later volcanism triggered by foundering IBC blobs like those observed in Case Ref (see Figure 3a, b for 1.99, 2.36 Gyr) does not occur (see the red line in Figure 8f and $L_{60}$ in Table 3).

The history of $\Delta R$ indicated for Case M5 in Figure 8e (the red line) is similar to that of Case Ref (the black line); $\Delta R$ increases for the first several hundred million years and then decreases with time. The amplitude of the increase is, however, considerably larger because a larger amount of magma is retained in the mantle owing to the lower permeability of this case. (In contrast to Case M5, we found a result closer to that of Case Ref at an elevated permeability of $M = 50$; Case M50 in Table 3 and Figure 8e, f.)

In summary, the lower permeability of Case M5 suppresses not only magma fingers but the percolation of magma through the lithosphere. The magma flux to the top of the mantle is therefore much smaller than that in Case Ref, and the mantle cools less efficiently (compare Figures 7 and 8 with Figures 3 and 4). The lack of magma fingers at this high retention number does not prevent partially molten plumes from forming to dominate mantle evolution, and the differences from Case Ref arise mainly from the suppressed percolation of magma through the lithosphere.

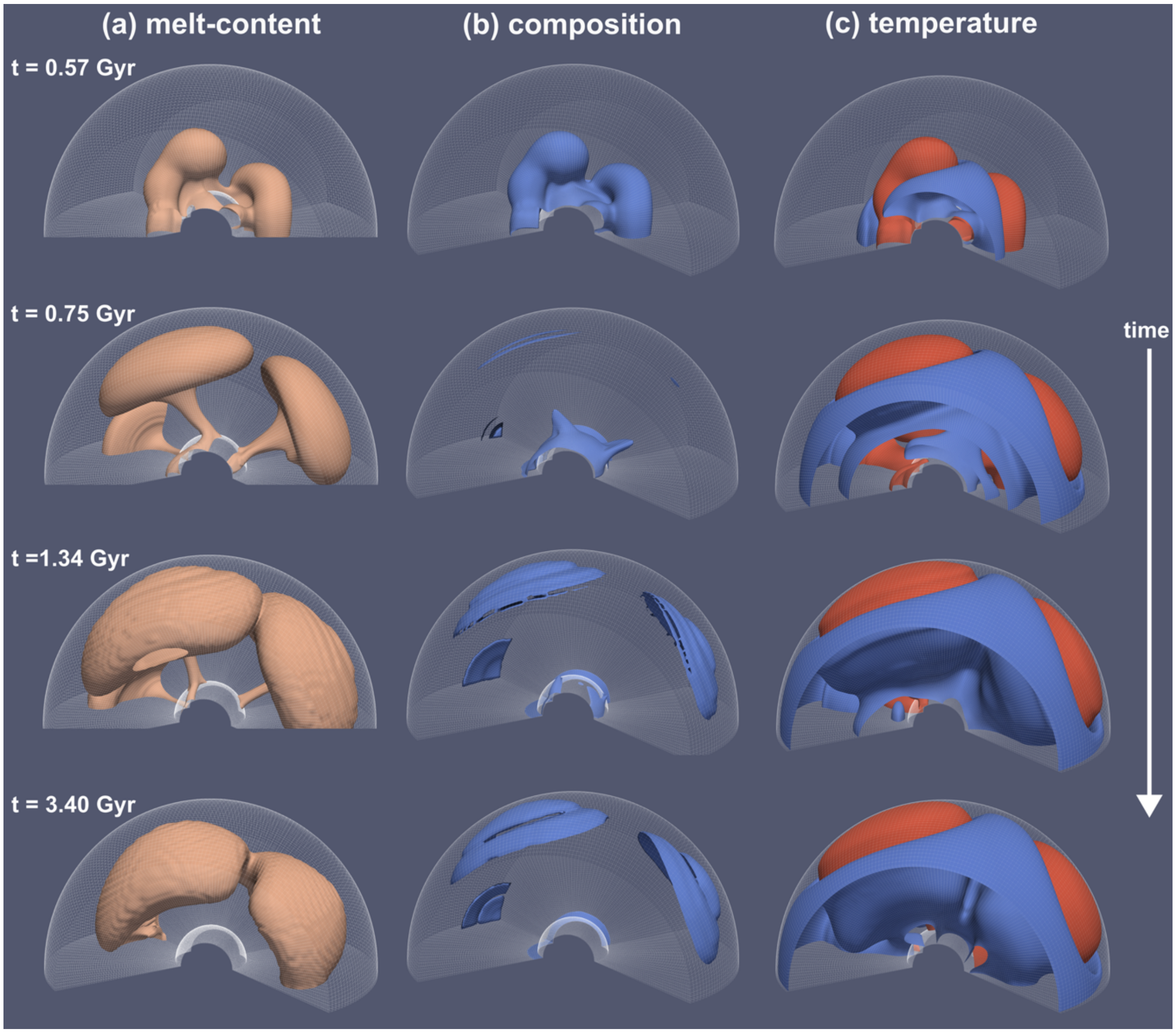

Figure 7. The same as Figure 3 but for Case M5 ($M = 5$ and other parameters fixed at their default values).

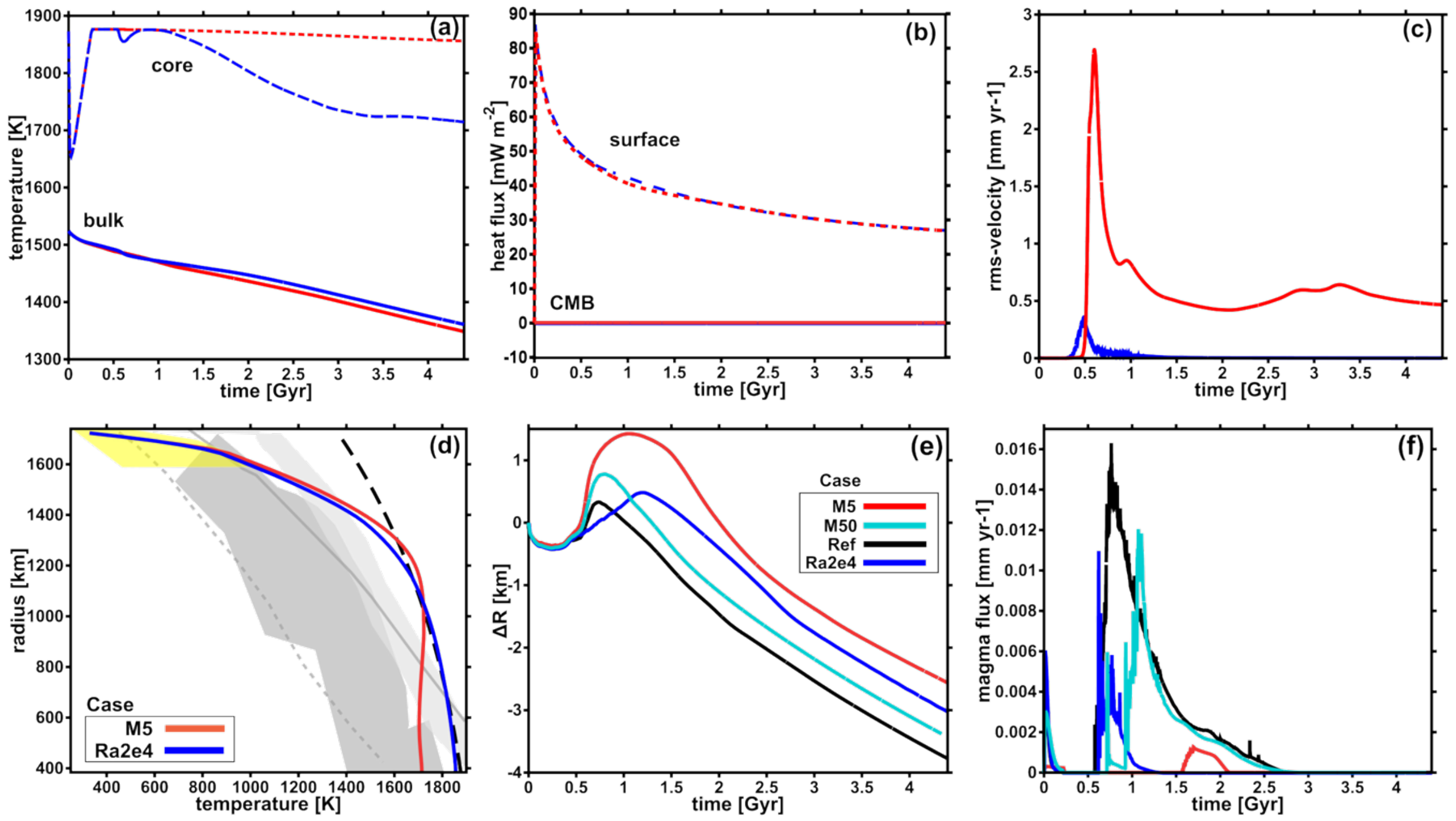

Figure 8. (a-d) The same as Figure 4 but for Case M5 (the red lines) and Case Ra2e4 (the blue lines). (e, f) Also shown are plots of radius change and magma flux at the depth level of 60 km against time calculated at various values of the reference permeability $M$ and the Rayleigh number $Ra$ (see Table 3).

#### 4.2.2 The Case Dominated by Magma Fingers (Case Ra2e4)

We next calculated Case Ra2e4 where the Rayleigh number is reduced to $Ra = 2.15 \times 10^4$, which is lower than the value assumed in Case Ref by a factor of 100 (Figure 9 and Movie S4). The permeability $M$ is still 100, the value for Case Ref, and hence the retention number of this case is about $10^4$, 100 times lower than that of Case Ref. At this low retention number, U et al. (2026a) predict that the melt percolation-dominated instability controls magma migration and that the MMUb feedback is too weak to drive partially molten plumes (Section 3).

Figure 9 shows that, indeed, only magma fingers develop in this case, and that partially molten plumes do not grow. Magma fingers grow upward to the uppermost mantle by around 0.67 Gyr and induce a peak in the plot of the magma flux at the depth level of 60 km (Figure 8f). Numerous small IBC blobs develop at the tips of the magma fingers. Because partially molten plumes do not develop, heat is not efficiently extracted from the deep mantle, and the mantle remains hot (see the blue lines in Figure 8d). The fingers laterally expand owing to melting by internal heating accordingly and partially melt most of the mantle (Figure 9a for 1.43 Gyr). The radial expansion $\Delta R$ reaches its peak at 1.2 Gyr due to this extensive melting (Figure 8e and Table 3). After that, the partially molten region shrinks monotonically with time owing to the conductive cooling from the surface boundary (Figure 9 for 3.11 Gyr; see the blue lines in Figure 8a-c). The magma fingers also cause a peak of the magma flux at around 0.6 Gyr (Figure 8f). In contrast to Case Ref, however, the flux declines rapidly after the peak and becomes zero by 1.1 Gyr, because partially molten plumes, which sustain the flux for billions of years in Case Ref, do not develop in this case. Note that the small IBC blobs formed by fingers reside in the stagnant lithosphere and do not founder, and hence they do not induce later magma ascent like that observed in Case Ref (Figure 9).

In short, magma segregates from the matrix so readily at this low Rayleigh number that partially molten plumes cannot grow. The volcanic activity therefore declines early, while the mantle, which loses little heat by magmatism, remains hot and cools only slowly (compare the blue lines in Figure 8a, d with the black line in Figure 4a).

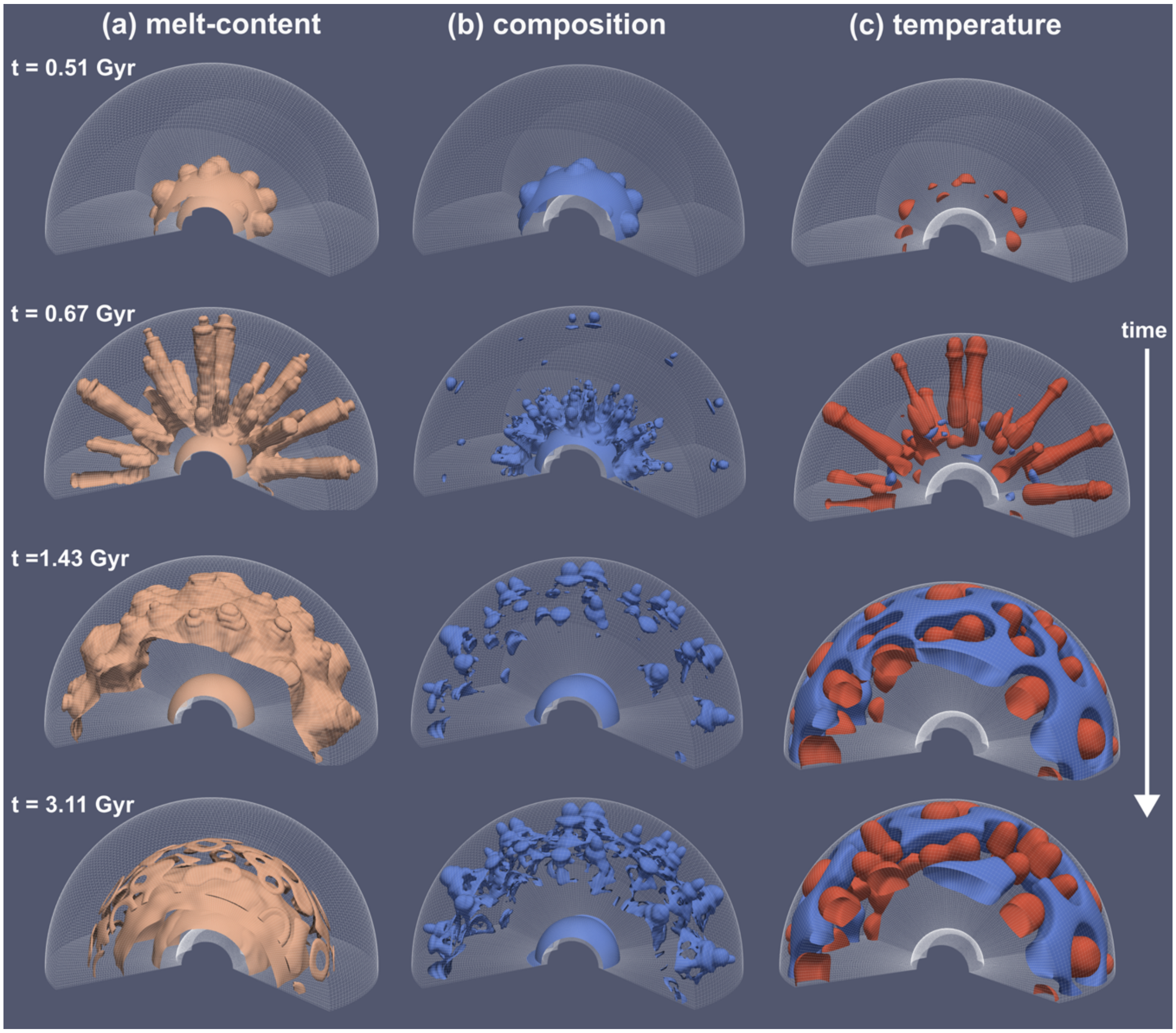

Figure 9. The same as Figure 3 but for Case Ra2e4 ($Ra = 2.15 \times 10^4$ and other parameters fixed at their default values).

### 4.2.3 The Effects of Density Inversion between Melts and Matrix

The solid phase is always denser than the IBC-rich melt phase in the cases described so far (see Eqs. (2) and (3) as well as Figure S3 in the Supporting Information). However, some studies suggest that IBC-rich magma can be denser than the coexisting matrix in the deep mantle of the Moon (e.g., Sakamaki et al., 2010; van Kan Parker et al., 2012; M. Xu et al., 2022). To see how this possible density inversion affects our numerical results, we calculated Case inv-vl, where $\Delta\rho_{\infty}/\rho_0 = 0.002$, $\Delta\rho_{\text{zero}}/\rho_0 = 0.35$, and $\lambda = 4.56$ GPa in Eq. (4) (see also Eq. (3)); the melt phase is denser than the coexisting matrix at radius $r$ less than around 800 km at these parameter values (Figure S3 in Supporting Information). Note that these parameter values also raise the solidus temperature through Eqs. (5) and (6); the solidus in Case inv-vl is higher than that in Case Ref by about 50 K in the uppermost mantle (the brown dashed curve in Figure 1a).

Interestingly, the overall feature of mantle evolution calculated in Case inv-vl (Figure 10 and Movies S5 and S6) is similar to that of Case M5 calculated at a lower permeability $M$ (Figure 7) rather than Case Ref that is calculated at the same permeability (Figures 3-6). For the first 600 Myr, magma is generated in the deep mantle and remains stable along the base of the mantle; magma fingers like those observed in Case Ref do not grow because magma migrates downward relative to the matrix. Instead, partially molten plumes grow directly from the partially molten region due to its Rayleigh-Taylor instability and carry the IBC-rich material at the base of the mantle to the uppermost mantle during the period of 0.71 to 1.85 Gyr (Figure 10 and Movie S5). The density inversion does not prevent the upwelling of these plumes, because melting always reduces the bulk density (Eqs. (1) to (3); see also U et al., 2023). Because the growth of partially molten plumes is delayed compared with that in Case Ref, the peaks of $\Delta R$ and the magma flux at the depth level of 60 km are also delayed: $\Delta R$ increases by 1.7 km to its peak at around 1.2 Gyr, and the magma flux peaks at around 2 Gyr (Figure 11a, b). When the plumes reach the uppermost mantle at around 1 Gyr, their heads spread laterally beneath the lithosphere (Figure 10 for 1.01 and 1.85 Gyr; see also Movie S5). The IBC-rich material carried by the plumes stays in these partially molten heads and does not founder (Figure 10b), and the later magma ascent triggered by foundering IBC blobs in Case Ref does not occur; the plume activity simply declines with time owing to cooling of the mantle (Figure 10 for 1.85 to 4.02 Gyr). The plume heads, however, remain partially molten even at 4.02 Gyr (Figure 10; Movies S5 and S6). $\Delta \mathrm{R}$ correspondingly decreases at a rate of about 1 km $\mathrm{Gyr}^{-1}$ after around 2.5 Gyr. Case inv-vl thus shares the main features of Case M5 (Section 4.2.1): no magma fingers; a few large plumes that rise directly from the basal partially molten layer; plume heads that spread laterally beneath the lithosphere and remain partially molten for billions of years; a magma flux to the depth level of 60 km that starts only after around 1.5 Gyr (Figures 8f and 11b); and IBC-rich blobs that do not founder.

Case inv-vl resembles Case M5 rather than Case Ref because the magma in the plume heads cannot easily percolate through the lithosphere, although the permeability is the same as in Case Ref. Two factors make the percolation harder than in Case Ref. First, the solidus is higher in Case inv-vl (Figure 1a). In Case Ref, pathways of magma grow above the plume heads into the lithosphere (see the arrows in Figure 3a for 0.70 Gyr and Movie S2 at around 0.9 Gyr). The higher solidus makes such pathways difficult to grow, and they are indeed poorly developed in Case inv-vl (Figure 10 for 1.85 Gyr; see also Movie S6). Second, the plumes grow later in Case inv-vl, as described above, and the lithosphere has become colder and thicker by the time the plumes reach its base (compare Movies S2 and S6). The melt therefore stays in the plume heads, which spread laterally beneath the lithosphere. The lithosphere is heated and thinned only above the centers of the heads, where the plume stems keep supplying melt and heat, and the magma eventually percolates to the depth level of 60 km there (Figure 10 for 1.85 Gyr; Movies S5 and S6). In Case M5, the low permeability itself suppresses the percolation. The similarity between the two cases thus shows that the ease of melt percolation through the lithosphere, in addition to the dynamics of partially molten plumes, strongly controls the overall evolution of the mantle (see also Section 5.3).

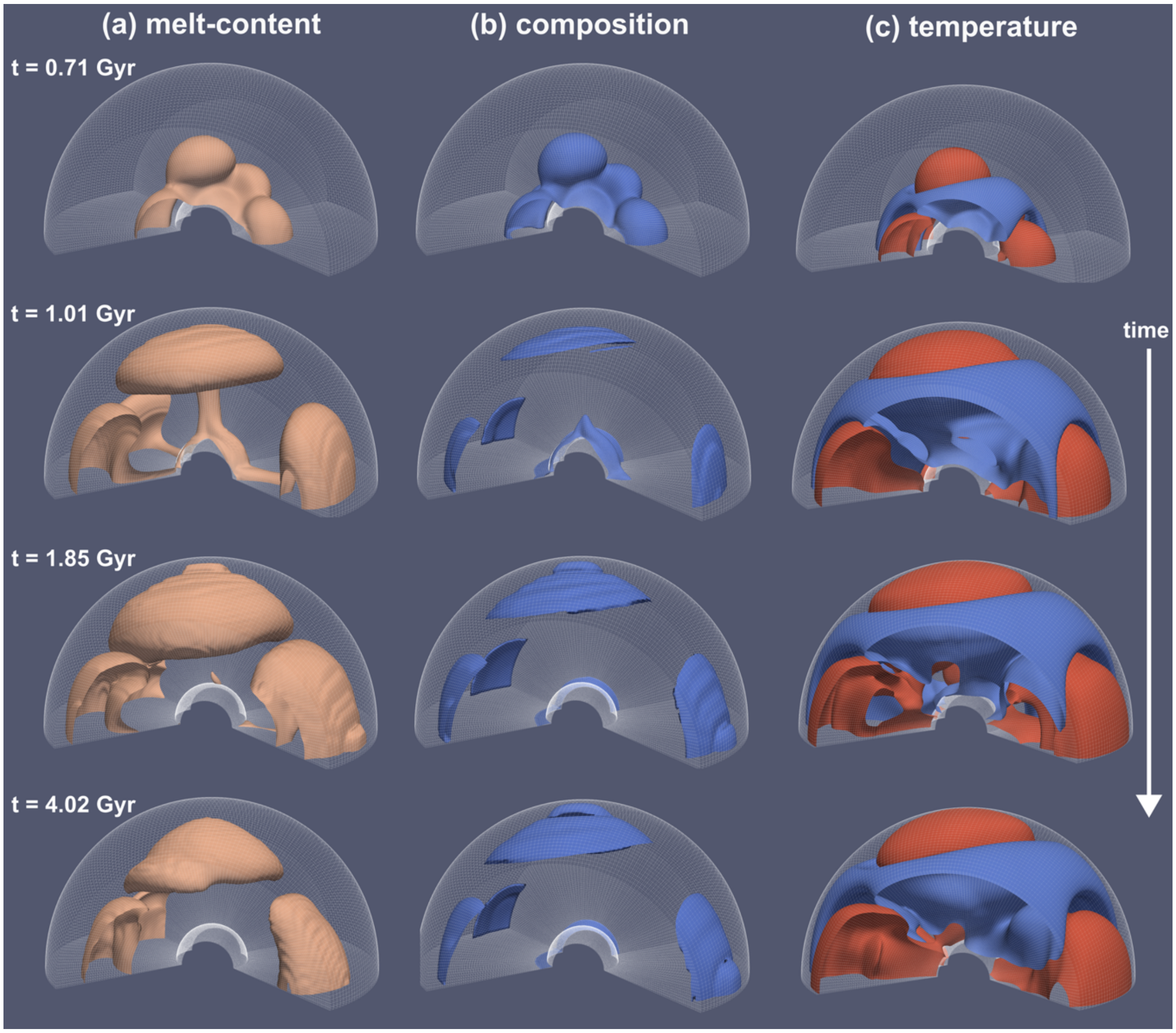

Figure 10. The same as Figure 3 but for Case inv-vl, where magma generated in the deep mantle is denser than the coexisting matrix.

## 4.3 Dependence on Other Model Parameters

We also varied the other model parameters listed in Table 2 to see how the results described above depend on them (Table 3; Figure 11; Figures S1 and S2 in Supporting Information).

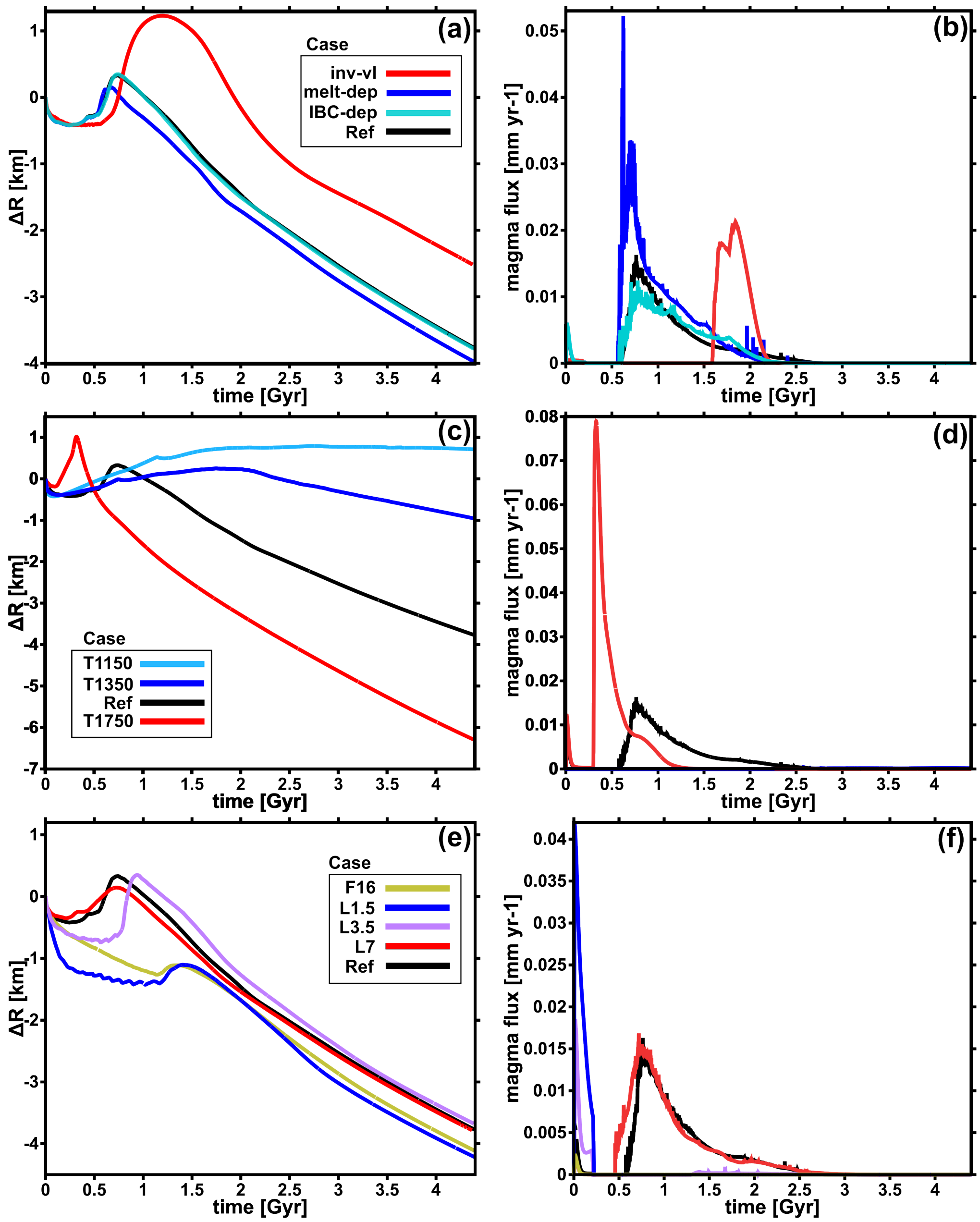

Figure 11. Plots of radius change and magma flux at the depth level of 60 km against time for the cases indicated in the figure: (a, b) the cases where the details of the rheology and possible density inversion between melts and matrix are considered; (c, d) the cases where the potential temperature $T_p$ of the mantle assumed in the initial condition is varied; (e, f) the cases where the thickness of the overturned layer $l^*$ and the initial ratio of HPE concentration in the crust to that in the mantle $F^*_{crst}$ are varied. The black lines show the reference case (Case Ref) presented in Figures 3-6.

#### 4.3.1 Melt- and IBC-Dependent Viscosity

So far, we have assumed that the viscosity of the mantle depends only on the temperature (see Eq. (9)) for simplicity. However, some earlier studies show that the viscosity depends on the contents of melt (e.g., Mei et al., 2002; Hirth & Kohlstedt, 2003; Scott & Kohlstedt, 2006) and the IBC component (e.g., Dygert et al., 2016; N. Zhang et al., 2017; H. Li et al., 2019). To consider the reduction of viscosity by melting and the IBC component, we calculated Cases melt-dep and IBC-dep: we assumed that the viscosity depends on the melt content $\phi$, the fraction of the olivine-rich material $F_{\mathrm{ol}} = (\xi_{\mathrm{b}} - \xi_{\mathrm{e}})/(1 - \xi_{\mathrm{e}})$ and temperature as:

$$\eta = [F_{\mathrm{ol}}\eta_0 + (1 - F_{\mathrm{ol}})\eta_{\mathrm{IBC}}]\exp\left[E_T(T_{\mathrm{ref}} - T) - \alpha_\phi \phi\right], \tag{28}$$

where $\alpha_\phi$ is the melt-weakening factor. Though some studies suggest $\alpha_\phi \approx 30$ (e.g., Hirth & Kohlstedt, 2003), we assumed $\alpha_\phi$ = 10 for numerical reasons; the viscosity reduction by melting is truncated at its value at $\phi = 0.4$. On the other hand, $\eta_{\mathrm{IBC}} = 7.5 \times 10^{18}$ Pa s is the viscosity of the IBC component (Dygert et al., 2016). In Case melt-dep, $F_{\mathrm{ol}}$ = 1 is assumed and the viscosity depends only on $\phi$, while in Case IBC-dep, $\alpha_\phi$ = 0 is assumed and the viscosity depends only on the content of the IBC component.

The overall features of mantle evolution calculated in Case melt-dep are the same as those of Case Ref. Magma generated in the deep mantle ascends first as magma fingers and then as partially molten plumes. The IBC blobs in the uppermost mantle formed by the plumes sink into the deep mantle, facilitating the counterflows and subsequent magma ascent. Because of the similarity to Case Ref, the history of radius change $\Delta R$ and the magma flux at the depth level of 60 km calculated for Case melt-dep (the blue lines in Figure 11a, b) are close to those of Case Ref (the black lines). The mantle-expands by 0.6 km from 0.3 to 0.7 Gyr due to the volume change by melting and then contracts until the end of the calculation (Figure 11a). The magma flux peaks at 0.6 Gyr due to the high activity of partially molten plumes at that time and then gradually declines to zero at 2.5 Gyr (Figure 11b). The peak in Case melt-dep is higher than that in Case Ref because the viscosity reduction by melting enhances the upwelling flow of partially molten plumes.

We found that the overall features of mantle evolution are not significantly affected by the IBC-dependent viscosity either (Case IBC-dep): both magma fingers and partially molten plumes developed as in Case Ref. The effects of the IBC-dependent viscosity are negligible in our study because the viscosity variation due to this dependence is only an order of magnitude at most for the variation in the content of the IBC component realized in the numerical models.

### 4.3.2 Dependence on the Initial Condition

We further explored the effects of the initial condition on the magma flux and radius change: the initial potential temperature $T_{\mathrm{p}}$, the thickness of the overturned layer $l^*$, and the initial ratio of HPE concentration in the crust to that in the mantle $F^*_{\mathrm{crst}}$ (Figure 11).

The initial potential temperature $T_{\mathrm{p}}$ affects the timing and, to some extent, the amplitude of the radial expansion and magma flux as shown in Figure 11c, d. A higher $T_{\mathrm{p}}$ leads to an earlier peak of the expansion and magma flux because partially molten regions develop earlier in the deep mantle to induce magma fingers and partially molten plumes. Magma generated in the deep mantle hardly reaches the uppermost mantle in the cases with $T_{\mathrm{p}} = 1350\ K$ or less (Figure 11d and Table 3) because most of the magma solidifies by cooling from the surrounding mantle before reaching the uppermost mantle. In these cases, the Moon contracts only slowly or hardly contracts in the later stage (Figure 11c and Table 3); when the initial temperature is low, the heat generated by internal heating exceeds the heat lost through the thick stagnant lid, and the temperature continues to rise for a few billion years.

The initial ratio of HPE concentration in the crust to that in the mantle $F^*_{\mathrm{crst}}$ and the thickness of the overturned layer $l^*$ also affect the radial expansion and magma flux (Figure 11e, f). A higher $F^*_{\mathrm{crst}}$ reduces radial expansion and causes the magma flux to drop to zero after 0.1 Gyr, because the mantle contains a smaller amount of HPEs and less magma is generated in the deep mantle (see the dark-yellow lines in Figure 11e, f). Similar results are obtained when the overturned layer is thick (see the blue lines in Figure 11e, f). Because the total amount of HPEs in the mantle is fixed (Eq. (24)), a thicker overturned layer spreads the HPEs over a larger volume and lowers their concentration near the CMB. The heating rate in the deep mantle then becomes too low to form a basal partially molten layer early, and less magma is generated there. In contrast, the histories of radius change and magma flux hardly change when the overturned layer is as thin as in Case Ref or thinner (compare the black and red lines for $1/l^* \geq 5.5$ in Figure 11e, f).

## 5 Discussion

Figure 3 illustrates how the mantle typically evolves in our 3-D regional spherical model of the lunar mantle. Strong internal heating induces a partially molten layer at the base of the mantle (Figure 3 for 0.4 Gyr), and the generated magma percolates upward through the matrix to induce magma fingers at the top of the layer by the melt percolation-dominated instability (Figure 3 for 0.56 Gyr). The magma fingers become thicker, while extending upward, and transform into partially molten plumes that ascend by the buoyancy of magma to generate more magma by decompression melting (that is, by the MMUb feedback; Figure 3 for 0.56 to 0.70 Gyr). The magma generated in the plumes eventually segregates and further percolates upward through the lithosphere to the surface; the flux of this percolating magma peaks during the period of 0.7 to 1 Gyr (Figure 6b). The growing magma fingers and partially molten plumes also cause radial expansion of the Moon that peaks during that period (Figure 6a). The radius and the magma flux then gradually decrease with time by cooling and solidification of the partially molten mantle (Figure 6). The magma flux, however, continues for billions of years owing to the continuing formation and growth of partially molten plumes caused by recycling of the magmatic products that are formed in the uppermost mantle by the earlier magmatism (see the white arrows in Figure 3 for 1.99 and 2.36 Gyr and Figure 6b). These plumes tend to grow at fixed locations and do not laterally migrate for billions of years. This typical behavior arises only when: (i) $R_t$ is not much lower than the threshold $R_{tb}$ defined by Eq. (26); (ii) the permeability is high enough for magma to segregate from plumes and to percolate through the lithosphere as suggested in Section 4.2.

### 5.1 3-D Modeling

A comparison of the 3-D model presented here with our earlier 2-D model (U et al., 2023) shows how the spatial dimension exerts control over the calculated history of mantle evolution. The overall features of mantle evolution for Case Ref summarized above are qualitatively the same as those of our 2-D models (U et al., 2023, 2025). Partially molten plumes calculated here, however, persistently ascend for billions of years at fixed locations (Figures 3 and 5) in contrast to those in our 2-D model that often migrate laterally over a significant distance. The 3-D geometry also makes the history of radial expansion/contraction and magma flux calculated here considerably different from that of our 2-D model: the early expansion in the 3-D model is only about a third of that in the 2-D model (Figure 6a), and the magma flux in the 3-D model has a lower peak than in the 2-D model, but it continues without interruption for more than 2 Gyr (Figure 6b).

In a 3-D geometry, plume positions become spatially stable because later upwellings exploit thermally softened conduits formed by earlier plumes. The heads of 3-D cylindrical plumes spread radially (see the arrows in Figure 12a), but perimeter cooling limits lateral coalescence, yielding isolated, persistent partially molten conduits. Because of the release of latent heat during crystallization within the conduits, their cooling is slower than that in the surrounding mantle, which promotes subsequent plume upwelling into the conduits. In contrast, in 2-D the heads of sheet-like, partially molten plumes spread laterally (see the arrows in Figure 12b; see also Figure 12c). With no out-of-plane direction available, their edge length effectively remains constant while the head elongates, and neighboring heads merge to form laterally extensive partially molten layers. There are, therefore, no thermally softened conduits for later upwellings. Our results are consistent with those of Guerrero et al. (2018), who showed that 2-D annular convection models tend to produce more laterally extended mantle plumes than 3-D spherical models, especially when the core is as small as in the Moon.

These persistent conduits are the reason why the magma flux in Figure 6b continues for a long time. In contrast to our 2-D models, the magma flux beneath the crust in the 3-D model continues without interruption for more than 2 Gyr (Figure 6b). This result may seem counterintuitive, because simple geometric reasoning would suggest that a 2-D sheet-like plume transports more magma at a time than a 3-D cylindrical plume. In the 2-D model, however, magma spreads laterally into broad layers that merge and cool efficiently, which leads to a higher peak but a shorter duration of the flux. In the 3-D model, later plumes repeatedly use the isolated conduits, which keeps the plume ascent continuous and sustains the magma flux over billions of years.

The difference in the shape of plume heads described above also controls the amplitude of the early expansion. The early expansion at around 0.7 Gyr occurs when the heads of partially molten plumes reach the base of the lithosphere and spread laterally. The circular heads in the 3-D model (the frame of 0.70 Gyr in Figure 3a; see also Figure 12a) cover a smaller area than the sheet-like heads in the 2-D model (Figure 12c). As a result, the early expansion in the 3-D model is only about a third of that in the 2-D model, and it is close to that in the 1-D spherically symmetric model (Figure 6a; U et al., 2022). The difference between the 3-D and 2-D models is much smaller for the rate of radial contraction during the past 1 Gyr (Figure 6a), but still significant: the contraction rate in the 3-D model is smaller than that in the spherically symmetric 1-D model where mantle convection does not occur. This is because the 1-D mantle cools less efficiently than the convecting 3-D mantle and remains partially molten as a whole even at present, so its solidification still contributes significantly to the contraction (see Figure 2a of U et al., 2022). In the 3-D model, in contrast, the mantle has mostly solidified by this time and only thermally contracts at present.

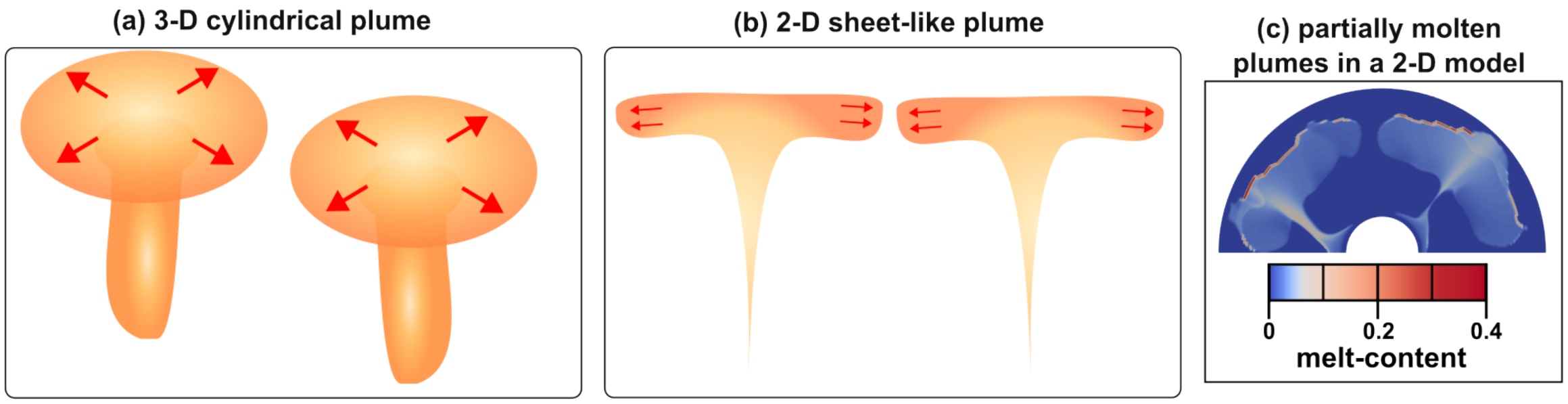

Figure 12. Schematic illustrations of partially molten plumes in (a) 3-D spherical geometry and (b) 2-D annular geometry. (c) A snapshot of melt-content at 1.12 Gyr in the reference case of U et al. (2023).

## 5.2 Upwelling Flows Driven by Melt Buoyancy

An important feature of our 3-D model is that the main driving force of mantle convection is the buoyancy of magma that is generated by decompression melting due to the mantle convection itself (e.g., Ogawa, 2018b; U et al., 2026a). Solid-state mantle convection driven by thermal buoyancy has often been considered in earlier 3-D models of lunar mantle convection (e.g., Laneuville et al., 2013; N. Zhang et al., 2013b). The difference in driving mechanism leads to several important differences in the calculated nature of upwelling flow of magma and matrix between our model and earlier models.

First, partially molten materials in our model reach the uppermost mantle and produce a significant flux of magma even at the depth level of 30 km (Figure 6b). Earlier researchers have numerically studied thermal convection in the lunar mantle to reproduce the mantle upwelling that causes lunar volcanism (e.g., Solomatov & Moresi, 1997; Zhong et al., 2000; Parmentier et al., 2002; W. Zhang et al., 2022). A linear stability analysis, however, shows that for the low Rayleigh number $Ra$, the strong temperature dependence of viscosity $E_T^*$, and the small core expected for the Moon, the mantle is below the threshold for thermal convection driven by basal heating (Yanagisawa et al., 2016); thermal convection may still be driven in the lunar mantle by internal heating, but it is probably weak. Indeed, Laneuville et al. (2013) found that a stagnant lithosphere as thick as about 35% of the mantle depth develops and that mantle upwelling occurs only as a return flow of cold plumes descending from the base of the lithosphere (see their Figure 4). In their model, mantle materials hot enough to cause volcanism reach the uppermost mantle only where a locally HPE-enriched region thins the lithosphere. Our model uses values of $Ra$ and $E_T^*$ comparable to theirs, but partially molten upwellings still reach the top of the mantle for two reasons. The upwellings are driven by their own melt buoyancy and actively rise through the mantle, and the latent heat of melting carried by the plumes keeps them hot while they ascend. Magma can thus make the upwelling flow that generates the magma itself substantially stronger than the upwelling flow of thermal convection.

Second, partially molten plumes in our model ascend stably at fixed locations and keep parts of the uppermost mantle partially molten for billions of years (Section 5.1). Earlier models have proposed two other ways to sustain localized volcanism. In some models of thermal convection in a 3-D spherical mantle, a single hot plume rises by its thermal buoyancy from a compositionally dense layer enriched in HPEs that extends globally (Zhong et al., 2000) or regionally (N. Zhang et al., 2017; W. Zhang et al., 2022, 2023) over the CMB. In these models, however, the stability of such a plume depends on model parameters such as the sensitivity of viscosity to temperature and the excess density of the basal layer, and it has yet to be examined whether these parameters allow a single thermal plume to grow in the Moon. In other models, a locally HPE-enriched region is placed at the top of the mantle and stays partially molten for more than 3 Gyr (e.g., Wieczorek & Phillips, 2000; Laneuville et al., 2014, 2018; Santangelo et al., 2025). These models, however, do not consider the extraction of HPEs from the enriched region by migrating magma. It is not clear whether such an enriched region stays partially molten even when magma migrates upward and extracts HPEs from it. Aside from these mechanisms, our results show that partially molten plumes driven by melt buoyancy can repeatedly rise through the conduits formed by earlier plumes and sustain the magma flux at fixed locations (Figure 6b and Section 5.4.4); consideration of this mechanism would be beneficial for future studies of localized lunar volcanism.

Third, partially molten plumes driven by melt buoyancy carry compositionally denser material from the deep mantle to the top of the mantle more efficiently than do solid plumes driven only by thermochemical buoyancy. Earlier studies have often addressed the fate of a convecting mantle placed on top of a basal layer that consists of compositionally denser material (e.g., Alley & Parmentier, 1998; Zhong et al., 2000; Le Bars & Davaille, 2002, 2004). When plumes in the convecting mantle are driven only by thermochemical buoyancy, whether the basal layer survives is controlled by the buoyancy ratio $B \equiv \Delta\rho_{\mathrm{comp}}/(\rho_0 \alpha_m \Delta T_{\mathrm{comp}})$, where $\Delta\rho_{\mathrm{comp}}$ and $\Delta T_{\mathrm{comp}}$ are the density and the temperature contrasts between the basal layer and the overlying mantle, respectively. When $B$ exceeds a critical value of order one, which depends on the viscosity and thickness ratios of the two layers, the basal layer remains stable; the plumes that rise in the mantle can then hardly entrain the dense material of the basal layer at $B \geq 1$ (e.g., Le Bars & Davaille, 2002, 2004; Lin & van Keken, 2006a, b; Heyn et al., 2018). In our models, in contrast, plumes driven by melt buoyancy extract and transport material from the compositionally dense IBC-rich basal layer even though $B$ is around 5 (see the frame of 0.70 Gyr in Figure 3); here, we estimated $B$ from $\Delta\rho_{\mathrm{comp}} = 170\ \mathrm{kg\ m^{-3}}$, the compositional density contrast across the IBC-rich layer, and $\Delta T_{\mathrm{comp}} = \mathrm{T_{Sc}} - T_p = 350$ K, where $T_{\mathrm{Sc}}$ is the solidus temperature at the CMB in Case Ref (Figure 1a, b). Melt buoyancy thus can overcome the barrier of compositional stratification and allow plumes to entrain the dense basal layer material even where purely thermochemical plumes cannot.

### 5.3 Percolation of Magma through the Lithosphere

The cases discussed in Section 4.2 show that the percolation of magma through the lithosphere strongly influences the mantle evolution. The importance of magma percolation through the lithosphere in mantle evolution has been already pointed out in Lourenço et al. (2016, 2018, 2020, 2026), who showed that the mode and efficiency of magmatism strongly control the thermal evolution and the tectonic regime of rocky planets. In their models, warm intrusive magmatism thins the lithosphere, whereas volcanic eruptions lead to a thick lithosphere that insulates the mantle. They concluded that the mantle therefore cools more efficiently when magma intrudes into the crust than when it erupts onto the surface (Lourenço et al., 2020). In our model, magma penetrates upward through the lithosphere only above partially molten plume heads whose locations are fixed (see Figure 3 and Movies S1 and S2), and the rest of the lithosphere behaves as a thick and cold stagnant lid. Our volcanism is, therefore, close to their volcanic eruptions. Yet, this volcanism efficiently cools the mantle by locally extracting magma from plume heads where most of the magma in the mantle resides (compare Case Ref with Cases M5 and inv-vl where segregation of magma from plume heads is not significant). Mantle cooling by localized magma segregation is so efficient, because sizable regions in deep mantle hot enough to bear magma always ascend as plumes; note that the buoyancy of magma is the dominant driving force of mantle upwelling in our model.

### 5.4 Implications for the Moon

#### 5.4.1 Volcanic Activity

The magma flux beneath the crust calculated in our reference case (Figure 6b) qualitatively accounts for the observed features of the lunar volcanic history indicated by histograms of the estimated ages of erupted basalts (e.g., Head et al., 2023). Apart from a short pulse at the beginning caused by the partial melting assumed in the initial condition (Section 4.1.3), the calculated magma flux is zero until about 0.5 Gyr (Figure 6b; see also Figure 3 for $t = 0.4$ Gyr); this early stage is likely to correspond to the period during which mare volcanism on the Moon was inactive (e.g., Hiesinger et al., 2000; Hurwitz et al., 2013). The peak of the flux at ~3.6 Gyr ago and its later gradual decline that continues until 2 Gyr ago observed in Figures 3a, 6b are consistent with the trend of the lunar mare volcanism that became active after around 4 Gyr ago, peaked between 3.8 and 3.5 Gyr ago, and then gradually declined over a period of billions of years (e.g., Hiesinger et al., 2003; Morota et al., 2009, 2011b; Whitten et al., 2011; Jin et al., 2025). The long tail of mare volcanism after the peak is likely a consequence of the partially molten conduits formed by earlier plumes, which persisted for billions of years and facilitated subsequent magma ascent (Sections 5.1 and 5.2). In the cases consistent with the observed histories of volcanism and radius change (Section 5.4.4), the total volume of magma that crossed upward the depth level of 30 km is about 1 to 10 km in global equivalent thickness (GET; $L_{30}$ in Table 3). This volume is comparable to or larger than the observed volume of lunar mare basalts, around 0.4 to 0.9 km GET, estimated from the average thickness and area of the maria (Head, 1976; Whitten & Head, 2015a; Sori et al., 2016; Broquet & Andrews-Hanna, 2024).

The persistent partially molten conduits observed in Figure 3 are also consistent with the localized nature of the lunar mare volcanism that continued for billions of years mainly in the PKT (e.g., Jolliff et al., 2000; Lawrence et al., 2000; Wieczorek & Phillips, 2000; Yang et al., 2025). The resurgence of volcanism at around 2 Gyr ago suggested by some studies (e.g., Morota et al., 2011b; Cho et al., 2012; Kato et al., 2017; Giguere et al., 2022; Tian et al., 2023) may be related to the magma ascent caused by counterflows of foundering IBC blobs, like the one that occurs at $t \approx 2$ Gyr in our model (Figure 3a, b). The magma generated by these counterflows is not enriched in HPEs, which does not contradict the HPE-depleted nature of the young basalts in the PKT (e.g., Che et al., 2021; Q. L. Li et al., 2021; Su et al., 2022; Wang et al., 2025).

Another implication of the plumes driven by melt buoyancy is that they can carry a significant amount of compositionally dense IBC-rich material from the base of the mantle to the source region of mare volcanism (Section 5.2). The compositional density contrast between the basal layer and the overlying mantle is 170 kg m$^{-3}$ in Case Ref ($1/l^* = 5.5$) and 270 kg m$^{-3}$ in Case L7 ($1/l^* = 7$) (see the black and red lines in Figure 1b). These values are comparable to those suggested by recent models of mantle overturn, typically 160 kg m$^{-3}$ (see Figure 6 in H. Li et al., 2019; see also Yu et al., 2019; Zhao et al., 2019; Maurice et al., 2024). Nevertheless, the plumes in Case Ref extract as much as 40% of the IBC-rich material from the basal layer assumed in the initial condition (Figures 1b and 3). The transport of the IBC-rich material to the source of mare volcanism at 4 to 3 Gyr ago is important for understanding the lunar mare basalts.

While our models reproduce the first-order temporal and spatial characteristics of lunar mare volcanism, they do not yet include several processes that are likely to be important. Tidal dissipation (Nimmo, 2025; W. Zhang & N. Zhang, 2025) and precession-driven core-mantle friction (Yu et al., 2024) may episodically enhance mantle melting and thereby modulate volcanic activity. The mechanical and thermal structure of the lithosphere further controls how efficiently basaltic magma percolates through the lithosphere to the surface (e.g., Morota et al., 2009; Whitten & Head, 2015b; Taguchi et al., 2017; Wilson & Head, 2017). Besides, a more general numerical framework for multiphase magma mixtures and their percolation through matrix would allow melt segregation, channelization, and eruption dynamics to be treated in a self-consistent manner (e.g., Spiegelman et al., 2001; Keller & Suckale, 2019; Wong & Keller, 2023). Our results represent a step toward an integrated model of lunar mantle evolution and mare volcanism that incorporates these additional processes.

#### 5.4.2 Radial Expansion/Contraction

The mantle evolution in our reference case also accounts for the radial expansion/contraction history of the Moon. In our reference case, the Moon expands globally by around 1 km for the first several hundred million years owing to the volume change associated with mantle melting (Figure 6a). The amplitude of early expansion is consistent with that of the Moon, 0.6-4.9 km, inferred from its gravity field in the literature (e.g., Andrews-Hanna et al., 2013; Sawada et al., 2016; Liang & Andrews-Hanna, 2022). After the radial expansion, the calculated mantle begins to contract radially at $t \approx$0.8 Gyr in our reference case (Figure 6a), which corresponds to the peak epoch of volcanism inferred from the magma flux shown in Figure 6b. The timing of the start of radial contraction is consistent with the onset of compressive tectonics on the Moon (Yue et al., 2017; Frueh et al., 2023). In the past one billion years, the mantle contracts at a rate of $0.9\ \mathrm{km\ Gyr^{-1}}$ in our model (Table 3), which is also consistent with the estimates obtained from observations of fault scarps on the Moon (e.g., Watters et al., 2010; Klimczak, 2015; Clark et al., 2017; van der Bogert et al., 2018; Nypaver et al., 2025). The early expansion and subsequent contraction in our models would help to understand the tectonic history of the Moon. Note that the 3-D geometry is essential for these quantitative estimates; our 2-D model predicts an early expansion about three times larger than that of the 3-D model and a faster contraction at present (see Section 5.1).

#### 5.4.3 Temperature Profile and Melt Distribution in the Mantle

It is interesting to compare the thermal structure of the mantle calculated at $t =$4.4 Gyr (present day) in our reference case with that inferred for the Moon. The depth profile of the minimum temperature at each depth at $t =$4.4 Gyr in our reference case is within the range of the mostly subsolidus temperature profiles inferred for the Moon from geophysical data (e.g., Sonett, 1982; Khan et al., 2006; Karato, 2013; Chang et al., 2026) throughout the entire depth range (Figure 4d). The maximum temperature, however, exceeds the solidus temperature in the depth range of 300-850 km (radius of 900-1450 km; see Figure 4d), and partially molten pockets are distributed there as shown in Figure 3 and Movie S1; most of the pockets are located at a depth of around 700 km. These partially molten pockets may account for deep moonquakes, which occur at fixed locations in the depth range of ~600-1200 km (e.g., Bulow et al., 2005, 2007; Nakamura, 2005; Frohlich & Nakamura, 2009); these deep moonquakes are suggested to be caused, or at least triggered, by tidal force (e.g., Cheng & Toksöz, 1978; Weber et al., 2009; Nahm et al., 2023). The fixed source of deep moonquakes may correspond to the gradually solidifying, yet stable, molten pockets depicted in Figure 3, because such pockets are expected to be mechanically weak and susceptible to tidal forcing. Indeed, the depths of the pockets are in the upper portion of the deep-moonquake focal range. The observation that the epicenters are primarily distributed on the nearside (Garcia et al., 2019; Nunn et al., 2020) where temperature is suggested to be higher than that on the farside (He et al., 2025; Park et al., 2025) is also consistent with the inference that the sources of deep moonquakes are partially molten pockets.

In contrast to the mid-mantle that hosts partially molten pockets, the deep mantle in our model is entirely solid at 4.4 Gyr. Our model therefore does not reproduce the laterally continuous melt layer that has been suggested to lie immediately above the CMB on the basis of the seismic low-velocity layer there (e.g., Latham et al., 1973; Nakamura et al., 1973; Weber et al., 2011; Matsumoto et al., 2015; Tan & Harada, 2021). The structure of the deep mantle would depend on some processes that are not included in our model, such as chemical reactions between the core and the mantle (e.g., Q. Xu et al., 2026). A more careful setting of the initial condition (Section 2.2) that takes these processes into account is necessary to understand the deep structure of the lunar mantle.

#### 5.4.4 The Initial State of the Lunar Interior

The initial conditions that lead to a mantle evolution model compatible with the observed lunar history of volcanism and radius change in our calculations may provide some clues for studies of the lunar magma ocean and subsequent mantle overturn. Among our models presented in Figures 8 and 11, the radius changes and magma fluxes calculated at $M \geq 50$, $Ra \approx 2 \times 10^6$, $T_\mathrm{p} \approx 1600$ K, ${F_\mathrm{crst}}^* < 16$, and $1/l^* > 3.5$ are consistent with the observations (see Section 4.3). When the potential temperature in the initial mid-mantle $T_\mathrm{p}$ is too low, 1350 K or less, the radial expansion occurs too late to reproduce the observed early expansion of the Moon, because it takes billions of years for internal heating to warm the mantle to the solidus (Figure 11c). At $T_\mathrm{p}$ as high as 1750 K, which makes most of the initial mantle partially molten, on the other hand, magma flux declines too early to reproduce the observed mare volcanism of the Moon that continued for billions of years (Figure 11d) as most of the IBC blobs are formed in the lithosphere and stay there; the blobs do not founder to cause counterflows and subsequent magma ascent. At a larger $F^*_\mathrm{crst}$, which means the mantle is depleted in HPEs, melting sufficient to drive the early expansion does not occur (Figure 11e). When the concentration of IBC material and HPEs toward the deep regions is gentle ($1/l^* \leq 3.5$), only a small amount of magma is generated in the deep mantle (Figure 11e, f); early expansion and peak of magma flux at around 4-3 Gyr ago like those observed for the Moon do not occur. In short, our simulations suggest that a substantial fraction of the mantle was solid, and a dense layer enriched in HPEs developed at the base of the mantle at the beginning of the history of the Moon. These suggestions may provide constraints on the evolution of lunar magma ocean (e.g., Boukaré et al., 2018; Maurice et al., 2020; Y. Zhang et al., 2025; Gu et al., 2026) and subsequent mantle overturn (e.g., Hess & Parmentier, 1995, 2001; Zhao et al., 2019; Scholpp & Dygert, 2026).

## 6 Concluding Remarks

To understand the history of volcanic activity and radial expansion/contraction of the Moon, we developed a 3-D regional spherical mantle convection model with magmatism, where magma generated by decompression melting of matrix percolates through the convecting matrix. The model shows that the buoyancy of magma drives the dynamics of the lunar mantle and sustains it for billions of years. In the early stage of lunar history, a basal partially molten layer is formed by internal heating (Figure 3 for 0.4 Gyr), and magma fingers start to grow from the top of the layer (Figure 3 for 0.56 Gyr). The magma fingers thicken as they extend upward and eventually evolve into partially molten plumes that rise owing to melt buoyancy to generate more magma by decompression melting (Figure 3 for 0.56 to 0.70 Gyr). Magma generated in these plumes by decompression melting eventually segregates and then percolates upward through the mantle to the surface, with the flux of this percolating magma reaching a maximum between 0.7 and 1 Gyr (Figure 6b). The growth of magma fingers and partially molten plumes extends the partially molten regions, and this extension causes a radial expansion of the Moon that peaks during the same interval (Figure 6a). Both the radius and the magma flux then gradually decrease as the partially molten mantle cools and solidifies (Figure 6). The magma flux, however, persists for billions of years, mainly because the partially molten plumes leave thermally softened conduits at fixed locations and later plumes repeatedly rise through these conduits (Figures 3 and 5): IBC blobs formed in the uppermost mantle by earlier magmatism founder and induce counterflows, which trigger renewed magma ascent through the conduits in the later stage (see the white arrows in Figure 3 for 1.99 and 2.36 Gyr and Figure 6b). The fixed locations of the conduits are a consequence of the 3-D geometry. This mantle evolution accounts for the peak of mare volcanism and the early radial expansion at around 4 to 3.5 Gyr ago, the localized volcanism that continued for billions of years, and the subsequent global contraction. The conditions required for the model to be consistent with the observed history suggest that the lunar mantle was mostly solid but still hot at the beginning of its history, with a dense layer enriched in HPEs and IBC material at its base. Because the present model is regional, fully global simulations that include global-scale structures (such as the lunar crustal dichotomy and the full geometry of the PKT) remain an important subject for future work.

## Appendix A: Non-Dimensional Forms of the Basic Equations

Here, we reduce the basic equations into their non-dimensional forms and present the definition of the non-dimensional parameters. The adopted length scale is $L = r_\mathrm{p} - r_\mathrm{c}$ ($r_\mathrm{p}$ and $r_\mathrm{c}$ are the radius of the surface and that of the CMB, respectively); the temperature scale is $\Lambda_0/C_p$; time scale is $L^2/\kappa$. The parameter values are shown in Tables 1 and 2.

The continuity equation is

$$\nabla^* \cdot \mathbf{U}^* = -\nabla^* \cdot [\phi(\mathbf{u}^* - \mathbf{U}^*)]\,, \tag{A1}$$

where the asterisks stand for normalized quantities.

The relative velocity $\mathbf{u}^* - \mathbf{U}^*$ is written as

$$\mathbf{u}^* - \mathbf{U}^* = Mg^* \frac{\phi^2}{\phi_0^3}\left\{\frac{\Delta V_\mathrm{l}}{V_0}[1 + \beta(1 - \xi_\mathrm{l})] - \beta(\xi_\mathrm{s} - \xi_\mathrm{l})\right\}\mathbf{e_r}\,, \tag{A2}$$

where

$$M \equiv \frac{k_{\phi_0}\rho_0 g_0 L}{\kappa \eta_\mathrm{melt}}\,. \tag{A3}$$

The momentum equation is

$$-\nabla^* p^* + Ra\Delta\rho^* \mathbf{e_r} + \nabla^* \cdot [\eta^*(\nabla^* \mathbf{U}^* + {}^t\nabla^* \mathbf{U}^*)\ ] \ = 0\,, \tag{A4}$$

where

$$\eta^* = \exp\left[E_T^*\left(T_{ref}^* - T^*\right)\right]. \tag{A5}$$

$\Delta\rho^* \equiv (\rho_0 - \rho)C_p/(\rho_0 \alpha_m \Lambda_0)$ is the dimensionless density deficit with respect to the reference density $\rho_0$. The Rayleigh number $Ra$ is defined as

$$Ra \equiv \frac{\rho_0 \alpha_{\mathrm{m}}\left(\Lambda_0/C_p\right) g_0 L^3}{\kappa \eta_0}\,. \tag{A6}$$

In the cases with the suffix -grad in Table 3, $Ra\,\Delta\rho^* \mathbf{e_r}$ in Eq. (A4) is replaced by $Ra\, g^* \Delta\rho^* \mathbf{e_r}$, where $g^* = g/g_0$ (see Section 2.1 and Text S1 in the Supporting Information).

The energy equation is

$$\frac{\partial h^*}{\partial t^*} + \nabla^* \cdot (h^* \mathbf{U}^*) = -\nabla^* \cdot [h_{\mathrm{l}}^* \phi(\mathbf{u}^* - \mathbf{U}^*)] - N^* g^* \frac{\Delta V_{\mathrm{l}}}{V_0} \phi u_r^* + \nabla^* \cdot (\kappa^* \nabla^* T^*) \\ + q^* + \nabla^* \cdot (\kappa_{\mathrm{edd}}^* \nabla^* h^*)\,, \tag{A7}$$

where $h_{\mathrm{l}}^* = h^*(\phi = 1)$ and

$$N^* \equiv \frac{g_0 L}{\Lambda_0}. \tag{A8}$$

## Appendix B: Linear Perturbation Analysis of Magma Migration in a Partially Molten Layer

Here, we summarize the linear perturbation analysis of U et al. (2026a) that leads to the two modes of magma migration described in Section 3 and to the threshold $R_{tb}$ in Eq. (26). (See also Spiegelman (1993a) and Hernlund et al. (2008b).) We consider a horizontal layer of a single-component partially molten material under the Boussinesq approximation in 2-D space. (Extension to 3-D space is straightforward.) The layer is a local model of the basal partially molten layer shown in Figure 3 for 0.4 Gyr. We take the horizontal coordinate $x$ and the vertical coordinate $z$ positive upward, and fix the gravitational acceleration at $g_0$. The viscosity of the matrix is constant at $\eta_0$, the eddy diffusivity $\kappa_{edd}$ in Eq. (13) is also constant, internal heating and thermal expansion are neglected, and the density difference between the solid and melt phases $\Delta\rho \equiv \rho_{\mathrm{s}} - \rho_l = \rho_0 \Delta V_l / V_0$ is a positive constant (Eqs. (2) and (3) with $\beta = 0$). Because the layer is entirely partially molten, the temperature is equal to the solidus temperature $T_{\mathrm{solidus}}$ given by Eqs. (5) and (6). We also neglected thermal diffusion in Eq. (13).

We non-dimensionalize the basic equations by the depth of the mantle $L$ for length, the percolation velocity $w_0 \equiv k_{\phi 0} \Delta\rho g_0 / \eta_{\mathrm{melt}}$ for velocity, $L/w_0$ for time, and $\Lambda_0 / C_p$ for temperature. The continuity equation (7) and Darcy's law (10) with Eq. (11) then become

$$\nabla^* \cdot [\phi \mathbf{u}^* + (1-\phi)\mathbf{U}^*] = 0, \tag{B1}$$

$$\phi(\mathbf{u}^* - \mathbf{U}^*) = \left(\frac{\phi}{\phi_0}\right)^3 \mathbf{e}_z, \tag{B2}$$

where $\mathbf{e}_z$ is the unit vector in the vertical direction. Eq. (B1) states that the total volume flux of melt and matrix is divergence-free. The momentum equation (8) becomes

$$-\nabla^* \hat{p}^* + R_t \phi \mathbf{e}_z + \nabla^{*2} \mathbf{U}^* = 0, \tag{B3}$$

where $\hat{p}^*$ is the non-dimensional dynamic pressure from which the term $\nabla^* \cdot \mathbf{U}^*$ is subtracted, and the retention number $R_t$ defined by Eq. (25) appears as the coefficient of the buoyancy of melt.

Because $T = T_{\mathrm{solidus}}$ in the layer, the enthalpy in Eq. (14) is written as $h^* = (T_0^* + \phi)(1+G)$ and that of the melt as $h_{\mathrm{l}}^* = (T_0^* + 1)(1+G)$. Within the layer, $G$ in Eq. (6) decreases linearly with height at the non-dimensional rate $S_d \equiv N^* \Delta V_{\mathrm{l}} / V_0$, where $N^*$ is defined by Eq. (A8). Substituting these expressions into the energy equation (13) and using Eq. (B1), we obtain an equation for the melt content

$$\frac{\partial \phi}{\partial t^*} + \nabla^* \cdot (\phi \mathbf{u}^*) = S\,[\phi u_z^* + (1-\phi) U_z^*] + \kappa_e^* \nabla^{*2} \phi, \tag{B4}$$

where $S \equiv T_0^* S_d / (1 + G_c)$ and $\kappa_e^*$ is the eddy diffusivity non-dimensionalized by $w_0 L$. Here, $G_c$ is the value of $G$ at the center of the layer; we assume that $S$ is a constant because the layer is thin (U et al., 2026a). The first term on the right-hand side of Eq. (B4) represents melting by decompression. The upward flux of melt and matrix $\phi u_z^* + (1-\phi) U_z^*$ carries the material to a lower pressure where the solidus is lower, and the enthalpy conservation requires that the material melts.

A steady solution to Eqs. (B1), (B2), and (B4) is a uniform melt content $\phi = \bar{\phi}$ with

$$\bar{U}_z^* = -\frac{\bar{\phi}^3}{\phi_0^3}, \qquad \bar{u}_z^* = \frac{\bar{\phi}^2}{\phi_0^3}(1-\bar{\phi}), \tag{B5}$$

where melt percolates upward and the matrix sinks to compensate for the melt flux. We decompose $\phi$, $\mathbf{U}^*$, and $\mathbf{u}^*$ into the steady parts and small perturbations as $\phi = \bar{\phi} + \delta\phi$, $\mathbf{U}^* = \bar{\mathbf{U}}^* + \delta\mathbf{U}^*$, and $\mathbf{u}^* = \bar{\mathbf{u}}^* + \delta\mathbf{u}^*$. The linearized forms of Eqs. (B1) to (B4) are

$$\delta\mathbf{u}^* = \delta\mathbf{U}^* + \frac{2\bar{\phi}}{\phi_0^3}\,\delta\phi\,\mathbf{e}_z, \tag{B6}$$

$$\nabla^* \cdot \delta \mathbf{U}^* = -\frac{3\bar{\phi}^2}{\phi_0^3}\frac{\partial \delta\phi}{\partial z^*}, \tag{B7}$$

$$-\nabla^* \delta \hat{p}^* + R_t\, \delta\phi\, \mathbf{e}_z + \nabla^{*2}\delta \mathbf{U}^* = 0, \tag{B8}$$

$$\frac{\partial \delta\phi}{\partial t^*} + V_z^* \frac{\partial \delta\phi}{\partial z^*} = S\left(\delta U_z^* + \frac{3\bar{\phi}^2}{\phi_0^3}\, \delta\phi\right) + \kappa_e^* \nabla^{*2}\delta\phi, \tag{B9}$$

where

$$V_z^* = \frac{3\bar{\phi}^2 - 4\bar{\phi}^3}{\phi_0^3}. \tag{B10}$$

Eq. (B7) shows that the matrix expands where the upward flux of percolating melt converges and contracts where it diverges. To solve Eqs. (B7) and (B8), we write the perturbation of matrix velocity in terms of a potential function $f^*$ and a stream function $\psi^*$ as $\delta U_x^* = -\partial f^*/\partial x^* + \partial \psi^*/\partial z^*$ and $\delta U_z^* = -\partial f^*/\partial z^* - \partial \psi^*/\partial x^*$. The former expresses the flow of matrix that compensates for the segregation of melt, and the latter expresses the circulation of matrix driven by the buoyancy of melt (Scott, 1988). Eqs. (B7) and (B8) reduce to

$$\nabla^{*2} f^* = \frac{3\bar{\phi}^2}{\phi_0^3}\frac{\partial \delta\phi}{\partial z^*}, \qquad \nabla^{*4}\psi^* = R_t \frac{\partial \delta\phi}{\partial x^*}. \tag{B11}$$

For a perturbation of the form

$$\delta\phi = A(t^*)\cos(k_x^* x^*)\exp(i k_z^* z^*), \tag{B12}$$

where $k_x^*$ and $k_z^*$ are the horizontal and vertical wavenumbers non-dimensionalized by $1/L$, Eq. (B11) gives

$$\delta U_z^* = \left[\frac{k_x^{*2} R_t}{(k_x^{*2} + k_z^{*2})^2} - \frac{3k_z^{*2}}{k_x^{*2} + k_z^{*2}}\frac{\bar{\phi}^2}{\phi_0^3}\right]\delta\phi. \tag{B13}$$

Substituting Eq. (B13) into Eq. (B9), we find that the perturbation grows as $A \propto \exp(t^*/\tau^*)$ while it travels upward at the velocity $V_z^*$, with the growth rate

$$\tau^{*-1} = S\,\frac{k_x^{*2}}{k_x^{*2} + k_z^{*2}}\left(\frac{R_t}{k_x^{*2} + k_z^{*2}} + \frac{3\bar{\phi}^2}{\phi_0^3}\right) - \kappa_e^*(k_x^{*2} + k_z^{*2}). \tag{B14}$$

Both terms in the parentheses of Eq. (B14) arise from decompression melting, but they represent different mechanisms. The first term, proportional to $R_t$, represents the MMUb feedback. Eq. (B13) shows that, when $R_t$ is large, the matrix ascends where the melt content is high as illustrated in Figure 2a. The ascent causes decompression melting and further increases the melt content there. The second term represents the melt percolation-dominated instability. When $R_t$ is small, the matrix descends where the melt content is high (Eq. (B13)) as illustrated in Figure 2b. The perturbation, however, still grows because the upward percolation of melt itself causes melting (the term proportional to $3\bar{\phi}^2/\phi_0^3$ in Eq. (B9)). This perturbation migrates upward at $V_z^*$ as a wave (Richter & Daly, 1989; Spiegelman, 1993a).

The two mechanisms contribute equally to the growth rate when

$$R_t = \frac{3\bar{\phi}^2}{\phi_0^3}(k_x^{*2} + k_z^{*2}) \tag{B15}$$

(Eq. (43) of U et al., 2026a). Since $k_x^* \approx k_z^*$ holds for the fastest growing mode as discussed in U et al. (2026a), the vertical velocity of matrix in Eq. (B13) changes its sign at around this value of $R_t$. For a perturbation whose horizontal and vertical sizes are both $d$, $k_x^* \approx k_z^* \approx 2\pi L/d$, and Eq. (B15) reduces to the threshold $R_{tb}$ in Eq. (26). The MMUb feedback dominates at $R_t \gg R_{tb}$, and the melt percolation-dominated instability dominates at $R_t \ll R_{tb}$. U et al. (2026a) confirmed by numerical simulations that the former generates partially molten plumes and the latter generates magma fingers when a solid layer overlies the partially molten layer, as described in Section 3.

## Acknowledgments

This work was supported by JST SPRING Grant JPMJSP2108, JSPS KAKENHI Grant JP22K03702, and JSPS KAKENHI Grant JP202412823 of Japan. This work was also supported by the Joint Usage/Research Center PRIUS at Ehime University, the Earth Simulator of Japan Agency for Marine-Earth Science and Technology (JAMSTEC), “Exploratory Challenge on Post-K Computer” (Elucidation of the Birth of Exoplanets [Second Earth] and the Environmental Variations of Planets in the Solar System), and “Program for Promoting Research on the Supercomputer Fugaku” (Toward a unified view of the universe: from large scale structures to planets). This work used computational resources of the supercomputer Fugaku provided by the RIKEN Center for Computational Science through the HPCI System Research Project (Project ID: hp230204, hp240219, and hp250226). Animations and some figures were drawn with ParaView, developed by Sandia National Laboratories, Kitware Inc., and Los Alamos National Laboratory.

## Data Availability Statement

The official version of the animations and Supporting Information as well as dataset will be made available if it has been accepted by JGR: Planets. However, the unofficial version of Figures S1-S7 can be accessed via the following links.

Movie S1: https://youtu.be/QM7yJ8Vr-bg

Movie S2: https://youtu.be/vD5XOUJNyyE

Movie S3: https://youtu.be/5aXt-WB8gs8

Movie S4: https://youtu.be/MICsU5fn7JQ

Movie S5: https://youtu.be/Gzph4ZQA_o0

Movie S6: https://youtu.be/IZ7nLvBRRQY

Movie S7: https://youtu.be/gs4Rq1qTGn8

## Conflict of Interest Disclosure

The authors declare there are no conflicts of interest for this manuscript.